\documentclass[review]{elsarticle}

\journal{Journal of Systems and Software}

\usepackage{hyperref}

\usepackage{url}

\usepackage{graphicx}
\usepackage{subfig}

\usepackage{multirow}
\usepackage{booktabs}
\usepackage{tabularx}
\usepackage{ragged2e}
\usepackage{caption}

\usepackage{amsmath}
\usepackage{amsfonts}
\usepackage{euscript}
\usepackage{amssymb}

\usepackage{algorithm}
\usepackage{algpseudocode}

\makeatletter
\def\BState{\State\hskip-\ALG@thistlm}
\makeatother

\algnewcommand\algorithmicforeach{\textbf{for each}}
\algdef{S}[FOR]{ForEach}[1]{\algorithmicforeach\ #1\ \algorithmicdo}

\usepackage{fixltx2e}

\usepackage{pdflscape}

\usepackage{changepage}   % for the adjustwidth environment

\usepackage{amsthm}
\theoremstyle{definition}
\newtheorem{definition}{Definition}

\usepackage{courier}
\usepackage{listings}
\begin{document}

\title{Architectural Stability Reasoning using Self-Awareness Principles: Case of Self-Adaptive Cloud Architectures}

%\author{Maria Salama, Rami Bahsoon, Rajkumar Buyya}
% Group authors per affiliation:
\author{Maria Salama\fnref{myfootnote}}
\ead{m.salama@cs.bham.ac.uk}
\author{Rami Bahsoon}
\address{School of Computer Science, University of Birmingham, UK}
\author{Rajkumar Buyya}
\address{Cloud Computing and Distributed Systems (CLOUDS) Lab, School of Computing and Information Systems, University of Melbourne}
\fntext[myfootnote]{Corresponding author}

\begin{abstract}
With the increased dependence on software, there is a pressing need for engineering long-lived software. As architectures have a profound effect on the life-span of the software and the provisioned quality of service, stable architectures are significant assets. Architectural stability tends to reflect the success of the system in supporting continuous changes without phasing-out. The \textit{behavioural} aspect of stability is essential for seamless operation, to continuously keep the provision of quality requirements stable and prevent architecture's drifting and phasing-out. In this paper, we present a framework for reasoning about stability during runtime, leveraging on self-awareness principles. Specifically, we employ runtime goals for managing stability goals, online learning for reasoning about stability on the long-run, and stochastic games for managing associated trade-offs. We evaluate the proposed work using the case of cloud architectures for its highly dynamics during runtime. The experimental results have shown the efficiency of self-awareness techniques in realising the expected behaviour stable during runtime operation.
\end{abstract}

\begin{keyword}
software architecture\sep 
cloud architecture\sep
stability\sep 
self-awareness\sep 
online learning\sep
stochastic games
\end{keyword}

%\linenumbers

\maketitle

\section{Introduction}
\label{sec_introduction}
Modern software systems are increasingly operating in highly open and dynamic and uncertain environments \cite{Chen2016}. Such challenges can have impact on the software life-time and the quality of service (QoS) provisioned. This growth, which is likely to continue into the foreseeable future, has motivated the need for long-lived software. An essential prerequisite for longevity of software systems is its capability to maintain service provision with expected qualities and accommodate changes in requirements and environment. 

An extensive literature survey \cite{Salama2018a} has revealed that the stability property has been considered at different levels (e.g. code, design, architecture levels) and with respect to several aspects (e.g. logical, structural, physical). This implies many different interpretations for considering stability as a software property. At the architecture level, stability has been viewed as the ability to endure with changes in requirements and the environment, while reducing the likelihood of architectural drifting and phasing-out, by avoiding ripple structural modifications (over two or more versions the software) \cite{Bahsoon2003b} \cite{Bahsoon2009}. That is an \textit{evolutionary perspective} in considering stability, i.e. evolving the system through a number of releases \cite{Kramer2007}. Meanwhile, dynamic changes, which occur while the system is in operation, require quick and dynamic adaptations during runtime \cite{Kramer2007}. This calls for an \textit{operational perspective} of stability that is fundamental for software architectures, to ensure seamless operation. 

As architectures have a profound effect on the life-span of the software and QoS provision \cite{Garlan2000} \cite{Garlan2015}, the architecture's behaviour tends to reflect the success of the system in constantly provisioning end-users' requirements, as well as supporting and tolerating continuous runtime changes \cite{Salama2017b}. We argue that architectural stability manifests itself as a software property necessary for the operation of software systems, their dependability and longevity over time. To leverage the capabilities of software systems, it is necessary to consider \textit{behavioural stability} to ensure that the architecture's intended behaviour is provisioned during runtime operation. This imposes new questions on how to take stability-aware adaptation decisions during runtime that can keep the architecture stable on the long-run. 

Even though some adaptation properties have been investigated by researchers in the self-adaptive software community, such as latency of adaptation (e.g. \cite{Camara2014c} \cite{Camara2015}), stability has not been tackled yet. On the other side, self-awareness has been recently employed in software engineering for reasoning and engineering better adaptations \cite{Elhabbash2018}. But self-awareness principles have not been employed for reasoning about runtime stability.

Achieving behavioural stability for long-living software calls for more intelligent reasoning about stability on the long-run. We propose reasoning about stability based on self-awareness principles. The principles of self-awareness are employed to enrich self-adaptive architectures with awareness capabilities, namely goal-, time and meta-self-awareness. We embed different computational intelligence techniques in self-awareness components for reasoning about stability.

\textbf{Contributions.} In more details, the main contributions are as follows.
\begin{itemize}
\item \textit{Goal-awareness for managing stability goals.}
With the typical key role of architectures in achieving quality requirements \cite{Kazman1994b} \cite{Sommerville2011} \cite{Barber2003} \cite{Ameller2013}, we can evidently agree that realising stability at the architecture level should be based on the quality requirements subject to stability \cite{Kazman1994b} \cite{Ameller2013} \cite{Angelopoulos2013}, where requirements are the key to long-term stability and sustainability \cite{Becker2016} \cite{Chitchyan2016}. We present runtime goals modelling for stability and implement algorithms for realising symbiotic relation between runtime goals model and self-awareness.

\item \textit{Time-awareness using online learning.}
Stability learning is essential for achieving stability on the long-term by learning from historical information. We propose a learning technique based on Q-learning, a reinforcement learning technique that can handle problems with stochastic transitions while learning how to act optimally in a controlled Markovian context \cite{Watkins1992} \cite{Dayan2006} \cite{Poole2010}. Time-awareness is, then, capable to take adaptation decisions converging towards stability by learning from historical information about adaptation actions and stability states.

\item \textit{Meta-self-awareness for managing trade-offs using model verification of stochastic games.}
Achieving a stable state for the architecture requires an explicit trade-offs management between different quality attributes, so that the adaptation process converges towards runtime goals given runtime uncertainty. We build a runtime approach for managing trade-offs based on automatic verification of stochastic multi-players games (SMGs) using PRISM-games 2.0 \cite{Chen2013b} \cite{Kwiatkowska2016}. The approach allows reasoning about possible adaptations for multiple attributes on the long-run.

\item \textit{Evaluation using the case of self-adaptive cloud architectures.}
We apply the proposed framework to the case of cloud architectures, where the continuous satisfaction and provision of quality requirements without SLA violations in the highly dynamic operating environment are challenging. The cloud architecture is modelled and simulated by extending \textit{CloudSim} \cite{Calheiros2011}. Our work is experimentally evaluated using the \textit{RUBiS} benchmark \cite{RUBiS} varying the number of requests proportionally according to the \textit{World Cup 1998} workload trend \cite{WorldCup98}. Experimental results have shown that the proposed approaches have proven feasibility in reasoning about stability during runtime.
\end{itemize}

\textbf{Organisation.} The rest of the paper is organised as follows. In section \ref{sec_background}, we describe relevant background. In section \ref{sec_stability}, we sketch the properties of architectural stability as a software property. Section \ref{sec_selfAwareness} elaborate the technical contributions on self-awareness techniques for stability reasoning. Section \ref{sec_evaluation} discusses experimental evaluation on the case of cloud architectures. We discuss the threats to validity of the proposed work and related work in section \ref{sec_threats} and \ref{sec_relatedWork} respectively. Section \ref{sec_conclusion} concludes the paper and indicates future work.

\section{Background}
\label{sec_background}
In this section, we introduce the main concepts (section \ref{sec_background_concepts}). Then, we present an overview on self-awareness (\ref{sec_background_self-awareness}).

 \subsection{Definitions of the Main Concepts}
\label{sec_background_concepts}

\paragraph{Software Architecture} 
The concept of software architecture has been defined in different ways under different contexts. In our work, we adopt the definition of the ISO/IEC/IEEE Standards that defines software architecture as the ``fundamental organisation of a system embodied in its components, their relationships to each other, and to the environment, and the principles guiding its design and evolution'' \cite{ISO2010Vocab}. This definition is in line with early definitions when the discipline has emerged \cite{Perry1992} \cite{Shaw1996} and with matured ones appearing later \cite{Bass2003}. Software architectures provide abstractions for representing the structure, behaviour and key properties of a software system \cite{Shaw1996}. They are described in terms of software components (computational elements), connectors (interaction elements), their configurations (specific compositions of components and connectors) and their relationship to the environment \cite{Medvidovic2000} \cite{Seo2008}. 

\paragraph{Software life cycle} 
The life cycle of a software system consists basically of the \textit{development} and \textit{operation} phases \cite{Avizienis2004}. The development phase includes all activities till the decision that the software is ready for operation to deliver service, such as requirements elicitation, conceptual design, architectural design, implementation and testing \cite{Avizienis2004}. The operation phase begins when the system is deployed, configured and put into operation to start delivering the actual service in the end-user's environment, cutover issues are resolved, and the product is launched \cite{Avizienis2004} \cite{ISO2010Vocab}. The former phase is known as \textit{initial development} or \textit{design-time}, and the latter is usually referred as \textit{runtime}. After the development and launch of the first functioning version, the software product enters to different cycles of maintenance and evolution stages till reaching the phase-out and close-down \cite{Rajlich2000} \cite{Avizienis2004} \cite{ISO2010Vocab}. During the maintenance stage, minor defects are repaired, while the system functionalities and capabilities are extended in major ways in the evolution stage \cite{Rajlich2000}.

\paragraph{Quality Attribute} 
The definition of a quality attribute we use is of the IEEE Standard for a Software Quality Metrics defining quality attribute as ``a characteristic of software, or a generic term applying to quality factors, quality sub-factors, or metric values'' \cite{IEEE1998Quality}. According to the same standard, a \textit{quality requirement} is defined as ``a requirement that a software attribute be present in software to satisfy a contract, standard, specification, or other formally imposed document'' \cite{IEEE1998Quality}.

\paragraph{Architecturally-significant requirements}
Generally, the architecture should fulfil the software requirements, both functional requirements (what the software has to do) and quality requirements (how well the software should perform) \cite{Witt1994} \cite{Gomaa2010}. Functional requirements are implemented by the individual components, while the quality requirements are highly dependent on the organisation and communication of these components \cite{Sommerville2011}. In the software architecture discipline, the architecturally-significant requirements are considered, as not all requirements have equal effect on the architecture \cite{Lianping2013}. Architecturally-significant requirements are a subset of technically challenging requirements, technically constraining and central to the system's purpose. These requirements have significant influence on the architecture design decisions, as they should be satisfied by the architecture \cite{Lianping2013}. \textit{Architecturally-significant functional requirements} may define the essence of the functional behaviour of the system \cite{Anish2014}, while \textit{architecturally-significant quality requirements} are often technical in nature, such as performance targets \cite{Kazman1994a} \cite{Ameller2013}. This special category of requirements, describing the key behaviours that the system should perform, plays a main role in making architectural decisions and has measurable effect on the software architecture. 

\paragraph{System Behaviour} 
The behaviour of a system is the ``observable activity of the system, measurable in terms of quantifiable effects on the environment whether arising from internal or external stimulus'' \cite{ISO2010Vocab}. This is determined by the state-changing operations the system can perform \cite{ISO2010Vocab}.

\paragraph{Self-adaptive software system} 
In general settings, \textit{to adapt} means ``to change a behaviour to conform to new circumstances'' \cite{Astrom1989}. A self-adaptive software ``evaluates its own behaviour and changes behaviour when the evaluation indicates that it is not accomplishing what the software is intended to do, or when better functionality or performance is possible'' \cite{Laddaga1997} \cite{Oreizy1999} \cite{Cheng2009a}. Intuitively, a self-adaptive system is one that has the capability of modifying its behaviour at runtime in response to changes in the dynamics of the environment (e.g. workload) and disturbances to achieve its goals (e.g. quality requirements) \cite{Meng2001}. Self-adaptive systems are composed of two sub-systems: (i) the managed system (i.e. the system to be controlled), and (ii) the adaptation controller (the managing system) \cite{Villegas2011}. The managed system structure could be either a non-modifiable structure or modifiable structure with/without reflection capabilities (e.g. reconfigurable software components architecture) \cite{Villegas2011}. The controller's structure is a variation of the MAPE-K loop \cite{Villegas2011}.

\subsection{Self-Awareness and Self-Expression}
\label{sec_background_self-awareness}
As self-adaptive software systems are increasingly becoming heterogeneous with dynamic requirements and complex trade-offs \cite{Nya2014}, engineering self-awareness and self-expression is an emerging trend in the design and operation of these systems. Inspired from psychology and cognitive science, the concept of self-awareness has been re-deduced in the context of software engineering to realise autonomic behaviour for software exhibiting these characteristics \cite{Lewis2011} \cite{Faniyi2014}, with the aim of improving the quality of adaptation and seamlessly managing these trade-offs.

The principles of self-awareness are employed to enrich self-adaptive architectures with awareness capabilities. As the architectures of such software exhibit complex trade-offs across multiple dimensions emerging internally and externally from the uncertainty of the operation environment, a self-aware architecture is designed in a fashion where adaptation and execution strategies for these concerns are dynamically analysed and managed at runtime using knowledge from awareness. 

\begin{figure}[!h]
\centering
\includegraphics[width=0.95\textwidth]{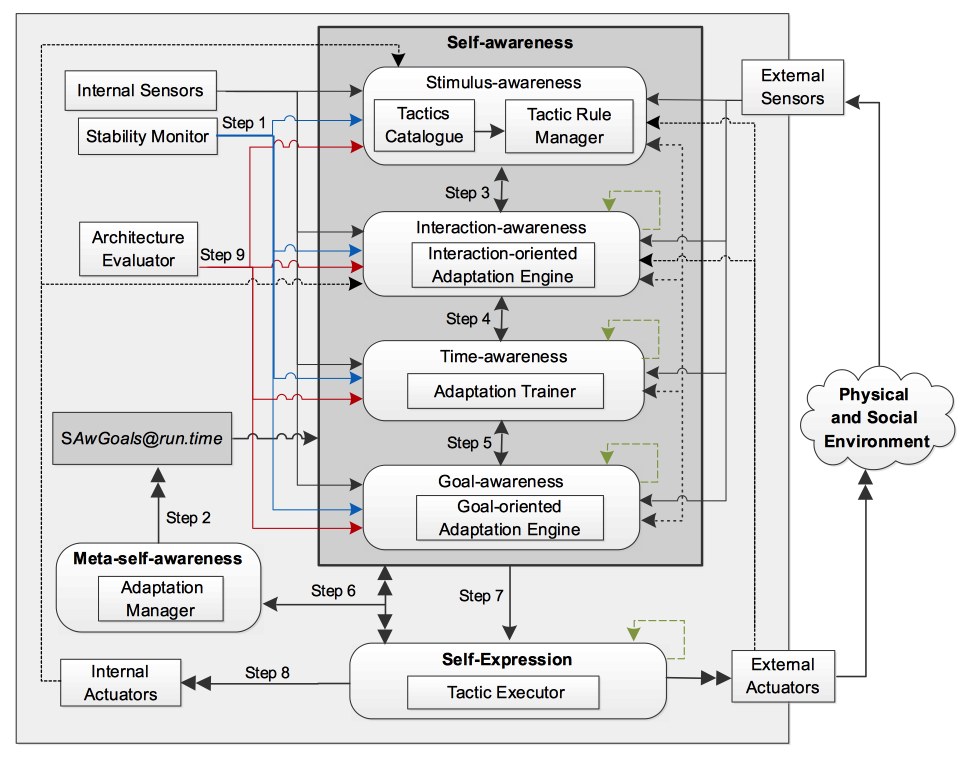}
\caption{Self-Awareness Framework}
\label{fig_metaPattern}
\end{figure}

The self-awareness framework is depicted in Figure \ref{fig_metaPattern}. A self-aware computational node is defined as a node that ``possesses information about its internal state and has sufficient knowledge of its environment to determine how it is perceived by other parts of the system'' \cite{Lewis2011} \cite{Faniyi2014}. A node is said to have \textit{self-expression} capability ``if it is able to assert its behaviours upon either itself or other nodes, this behaviour is based upon a nodes sense of its personality'' \cite{Parsons2011}. Different levels of self-awareness, called capabilities, were identified to better assist the self-adaptive process \cite{Parsons2011} \cite{Faniyi2014}: 
\begin{itemize}
	\item \textit{Stimulus-awareness}: a computing node is stimulus-aware when having knowledge of stimuli, enabling the system's ability to adapt to events. This level is a prerequisite for all other levels of self-awareness. 
	\item \textit{Goal-awareness}: if having knowledge of current goals, objectives, preferences and constraints, in such a way that it can reason about it. 
	\item \textit{Interaction-awareness}: when the node's own actions form part of interactions with other nodes and the environment. 
	\item \textit{Time-awareness}: when having knowledge of historical information and/or future phenomena. 
	\item \textit{Meta-self-awareness}: the most advanced of the self-awareness levels, which is awareness of own self-awareness capabilities.
\end{itemize}

Reviewing self-awareness in software engineering \cite{Elhabbash2018}, we found it has been employed in software systems for reasoning and engineering better adaptations with guaranteed functionalities and quality of service during runtime (e.g. \cite{Sterritt2005}, \cite{Lewis2011}), online service composition without human intervention \cite{Elhabbash2014}, realising intelligent software systems with sophisticated abilities \cite{Vassev2012}, and dealing with real-world situations and dynamic environments of modern software systems to respond to such fluctuating environment and associated uncertainty \cite{Becker2012} \cite{Chen2015}. Yet, behavioural stability has been explored in self-adaptation using self-awareness principles.

\section{Architectural Stability}
\label{sec_stability}
Generally, the notion of ``stability'' refers to the resistance to change and the tendency to recover from perturbations. The condition of being stable, thus, implies that certain properties of interest do not (very often) change relative to other things that are dynamically changing. As a software quality property, stability is defined in the ISO/IEC 9126 standards for software quality model \cite{ISO2000Quality} as one of the sub-characteristics of the maintainability characteristic of the software \textemdash along with analysability, changeability and testability \textemdash as ``the capability of the software product to avoid unexpected effects from modifications of the software'' \cite{ISO2000Quality}. For general application purposes, the standard does not determine specific features or aspects for stability \cite{Dobrica2002}. 

Reviewing the state-of-the-art in software engineering \cite{Salama2018a}, we have found that stability has been considered at different levels, i.e. at the code level (e.g. \cite{Yau1980}), requirements (e.g. \cite{Bush2003}), design (\cite{Yau1985} \cite{Fayad2001} \cite{Elish2005} \cite{Kelly2006}) and the architecture level (\cite{Jazayeri2002} \cite{Bahsoon2003b} \cite{Tonu2006} \cite{Molesini2010}). At each level, stability has been considered in relation to several aspects from different perspectives, and thus interpreted in many ways according to the perspective of consideration. For instance, stability at the code level has been interpreted as ``the resistance to the potential ripple effect that the program would have when it is modified'' \cite{Yau1980}, that is considering the \textit{logical} and performance (i.e. \textit{behavioural}) aspects of stability from the \textit{maintenance} perspective. Design stability has been referred to ``the extent to which the structure of the design is preserved throughout the evolution of the software from one release to the next'' \cite{Yau1985}, where the \textit{logical} and \textit{structural} aspects of stability are considered from \textit{evolutionary} perspective.

Architectural stability has been considered in terms of ripple structural modifications over two or more versions of the software, as a \textit{structural} aspect with respect to architecturally-relevant changes carried from \textit{evolutionary} (\cite{Jazayeri2002} \cite{Bahsoon2003b}) and maintenance perspectives (\cite{Molesini2010}). This has been referred to the extent to which the architecture's structure is capable to accommodate the evolutionary changes without re-designing the architecture or making ripple modifications \cite{Jazayeri2002} \cite{Bahsoon2003b}. Among different perspectives, the \textit{structural} aspect of stability is the one mostly considered at the architecture level.

Considering runtime dynamics of software systems, the structure and the behaviour of the software may be affected when adaptations are taking place during runtime \cite{Chomchumpol2015}. In this context, we distinguish between the \textit{structural} and \textit{behavioural} aspects of stability. We also posit that an \textit{operational} perspective (for the runtime operation of the software) is essential, different from the \textit{evolutionary} perspective (over two or more versions of the software). The stability meaning, we are seeking, can be regarded at the \textit{architecture} level considering the \textit{behavioural} aspect from an \textit{operational} perspective. As such, we define stability as \textit{the ability of the architecture's behaviour to maintain a fixed level of operation (or recover from operational perturbations) within specified tolerances under varying external conditions}. A stable architecture from the \textit{operational} perspective is the one capable to continuously fulfil the architecturally-significant quality requirements during runtime, where the architecture can return to the equilibrium state, following a perturbation due to changes in quality requirements, workload patterns or in the operational environment. Conversely, an unstable architecture is one that, when perturbed from equilibrium, will show deviation from the expected behaviour. So, stability of the architecture is essential to examine the behaviour with time following a perturbation during runtime. 
 
Given the runtime dynamics, self-adaptive architectures need to keep the QoS provision stable, as well as maintain the stability as an adaptation property \cite{Villegas2011} \cite{Villegas2017} \cite{Salama2017b}. The former form of stability corresponds to the quality attributes desired to be delivered as adaptation goals and how constantly they are delivered without SLA violations. The latter form of stability is the degree in which the adaptation process converges towards the adaptation objective. Both aspects, mapped to each other, are used to evaluate the quality of adaptation \cite{Villegas2011} \cite{Villegas2017} \cite{Salama2017b}.

\section{A Self-Awareness Assisted Framework for Reasoning about Architectural Stability}
\label{sec_selfAwareness}
We employ the different awareness capabilities for reasoning about architectural stability. The goal-awareness embeds a runtime goals modelling for managing stability goals and realising the symbiotic relation between the self-awareness component and goals during runtime (discussed in section \ref{sec_selfAwareness_goal}). The time-awareness implements an online learning algorithm to assist in making adaptation decisions leading to stability using historical information (discussed in section \ref{sec_selfAwareness_time}). The meta-self-awareness is assisted by probabilistic game-theoretic approach for managing trade-offs between different stability goals (discussed in section \ref{sec_selfAwareness_meta}). 

\subsection{Goal-Awareness for Managing Stability Goals}
\label{sec_selfAwareness_goal}
We propose the \textit{SAwGoals@run.time} component for managing stability goals (as illustrated in Figure \ref{fig_metaPattern}). As runtime goals drive the architecture in reasoning about adaptation during runtime \cite{Sawyer2010}, \textit{SAwGoals@run.time} extends the GORE model to suit the needs of self-awareness capabilities and stability requirements. The objectives of the proposed modelling are: (i) fine-grained dynamic knowledge representation of stability goals to enable efficient use of the different levels of self-awareness, (ii) monitoring the satisfaction of stability goals and the performance of tactics, (iii) better informed decision of the optimal tactic for realising architectural stability, and (iv) continuous accumulation of historical information to update the knowledge for future learning using time-awareness.

We refine the Runtime Goal Models with fine-grained dynamic knowledge representation that reflects self-awareness needs for new attributes of the goals, operationalisation, tracing down to architecture and runtime satisfaction measures. Specifically, additional runtime behavioural details relevant to different levels of self-awareness are integrated, such as node information for interaction-awareness, and trace history for time-awareness, as well as information about the execution environment in different time instances. Operationalisation of stability attributes is realised by self-expression, through runtime tactics which are defined within the proposed model. The model would better operate in the presence of historical information about the ability of operationalisation decisions. In the case of instantiation, it is imperative that the designer consider what-if analysis, simulation or scenarios to test the suitability of the choice. Models which rely on decision-making under uncertainty can also be sensible to employ. Given relevant information about goals and the operating environment, conflict management between goals during runtime is handled by meta-self-awareness capabilities.

\subsubsection{Runtime Goals Knowledge Representation}
Runtime goals in \textit{SAwGoals@run.time} are defined along with an execution trace and traced to runtime tactics for operationalisation. A \textbf{Runtime Goal} (e.g. performance) $ G \in \mathcal{G} $, where $ \mathcal{G} $ is the set of goals in a self-aware and self-expressive node. A goal is defined by the following attributes:

\begin{itemize}
\item \textit{Unique identifier} $ id $ of the goal $ G $.

\item \textit{Definition}. formally and informally defining the goal and its satisfaction in an absolute sense. 

\item \textit{Node identifier} $ N $, the unique identifier of the self-aware node responsible for realising the goal.

\item \textit{Weight} $ w $ to consider the priority of the goal.

\item \textit{Metric} $ M $ a measurable unit (e.g. response time measured in milliseconds) that can be used to measure the satisfaction of the goal while the system is running. 

\item \textit{Objective Functions} $ f(G) $ defines the measures for assessing levels of goal satisfaction with respect to values defined in SLAs of different end-users (e.g. objective functions for performance are response time 15 ms and 25 ms for dedicated and shared clients).

\item \textit{Set of tactics} $ T(G) \in \mathcal{T} $ to be used in case of violation of the goal. The goal semantic is the set of system behaviours, i.e. runtime tactics, that satisfy the goal's formal definition. 
\end{itemize}

A \textbf{Runtime Tactic} $ T \in \mathcal{T} $ (e.g. vertical scaling) is defined as follows:
\begin{itemize}
\item \textit{Unique identifier} $ id $ of the tactic $ T $.

\item \textit{Definition} includes description and informal definition for when to apply the tactic and how to execute it.

\item \textit{Object} in the architecture in which the tactic is executed (e.g. VMs).

\item \textit{Pre-condition} defines the current condition of the operating environment in which the tactic could be applied.

\item \textit{Limits} defines the minimum and maximum limits of the architecture for executing the tactic (e.g. the maximum number of servers).

\item \textit{Functionality} defines how the tactic should be executed.

\item \textit{Post-condition}. This characterises the state of the operating environment after applying the tactic. 

\item \textit{Variantions of the tactics} includes different forms or possible configurations for applying the tactic (e.g. earliest deadline first scheduling, least slack time scheduling).
\end{itemize}

A \textbf{Runtime Goal Instance} $ G(n, t_i) $ is an instance of the runtime goal $ G $ in the self-aware node $ n $ at a certain time instance $ t_i $, and is defined as follows:
\begin{itemize}
\item \textit{Client} $ c $ issuing the service request $ r $.

\item \textit{Objective function} denotes the quality value defined in the SLA of the client $ c $.

\item \textit{Tactic} $ T $ and its configuration executed as an adaptation action to satisfy the goal.

\item \textit{Actual value} $ v $ denotes the degree of satisfaction achieved after the execution of the tactic $ T $ that is measured by the Architecture Evaluator.

\item \textit{Set of environment runtime goals} $ G_e $, that are the goals from other self-aware nodes $ n_x $ running at the same time instance $ t_i $ with which the node $ n $ is interacting, where $ G_e = \{ G_1(n_1, t_i), G_2( n_2, t_i), ..., G_x(n_x, t_i) \} $.

\item \textit{Set of environment runtime tactics} $ T_e $, that are the tactics taking place at the same time instance $ t_i $ in the environment, where $ T_e = \{ T_1, T_2, ..., T_x \} \mbox{ for } \forall \mbox{ } G \in G_e $.
\end{itemize}

For each goal $ G $, \textit{change tuples} are created at different time instances $ t_i $ to form the history of this goal $ \mathsf{H}(G) $ for keeping record of the goal satisfaction and related tactics performance over time. This history shall be used by time-awareness to reason about adaptation actions in the future.

\subsubsection{Algorithms for Realising Symbiotic Relation Between Runtime Goals and Self-awareness}
\begin{figure}[!h]
\centering
\includegraphics[width=0.75\textwidth]{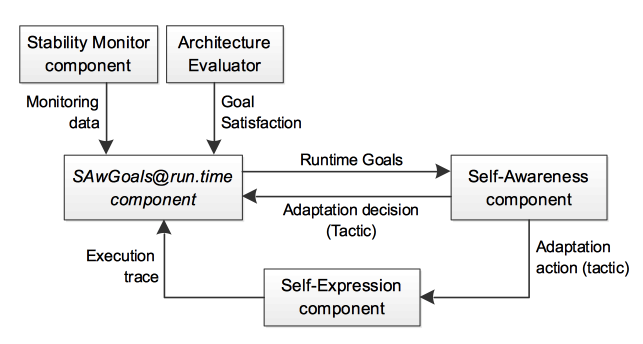}
\caption{Symbiotic relation between Runtime Goals and Self-awareness}
\label{fig_symbioticRelation}
\end{figure}

As end-users' requirements change during runtime, there is a need to maintain the synchronisation between the goals model and the architecture \cite{Sawyer2010}. We envision enriching the proposed architecture patterns and goals modelling by incorporating the symbiotic relation between runtime goals and self-awareness capabilities. The symbiotic relation promises more optimal adaptations and better-informed trade-offs management decisions. It aims to keep the runtime goal model ``live'' and up-to-date, reflecting on the extent to which adaptation decisions satisfied the goal(s). The symbiotic relation, illustrated in Figure \ref{fig_symbioticRelation}, is realised during runtime as follows.

\begin{enumerate}
\item Goals are defined and modelled in the \textit{SAwGoals@run.time} component, with fine-grained knowledge representation relevant to the different levels of awareness. 

\item Having goals information fed to the self-awareness component, a better informed adaptation decision would be taken based on the learning of time-awareness and the runtime environment of interaction-awareness capabilities.

\item The selected tactic is executed by the self-expression component. 

\item The execution trace is, then, fed back to the goals model to be kept in the log of the goal history. 

\item The goal satisfaction is evaluated by the Architecture Evaluator component to be logged in the goal history. 

\item The goal history is used, in turn, by time-awareness at the next time instance when selecting the appropriate tactic.
\end{enumerate}

To realise the symbiotic relation, we provide algorithms to process the Runtime Goal Instance (Algorithm \ref{alg_runtimeGoalInstance}) and construct the Goal History (Algorithm \ref{alg_goalHistory}). 

\textbf{Algorithm 1: Processing Runtime Goal.} This algorithm is launched to process the Runtime Goal Instance $ G(n, t_i) $ at time instance $ t_i $. 

\begin{center}
\begin{minipage}{0.75\linewidth}
\footnotesize
\begin{algorithm}[H]
\caption{Process Runtime Goal}
\label{alg_runtimeGoalInstance}
\begin{algorithmic}[1]
\Procedure {ProcessGoal}{$ G_i = (G_{id}, N_{id}, t_i) $} 
	\State get ObjectiveFunction(client $ c $) 
\BState \emph{QoSMonitor}:
	\State get MonitoringData($ G $)
\BState \emph{Self-awarenessComp}:
	\If {violation($ G $)} 
		\State Identify set of possible tactics $ T(G) $
		\If {TimeAwareness is enabled}
			\State get goal hisotry $ \mathsf{H}(G) $
		\EndIf
		\State select tactic $ T_x \in T(G) $
\BState \emph{Self-expressionComp }:
		\State execute tactic $ T_x $ 
		\State get ExecutionTrace $ \tau(G_i) $ 
\BState \emph{ArchitectureEvaluator}:
		\State get GoalSatisfaction $ v(G) $ 
	\EndIf
\EndProcedure
\end{algorithmic}
\end{algorithm}
\end{minipage}
\par
\end{center}

\textbf{Algorithm 2: Constructing Goal History.} 
This algorithm constructs a change tuple for the goal $ G $ at each time instance $ t_i $. Each change tuple records a log of the objective function, goals from the environment, set of tactics executed in the environment, the tactic executed, the execution trace and the goal satisfaction measure. These change tuples would form the goal history over the different time instances.

\begin{center}
\begin{minipage}{0.87\linewidth}
\footnotesize
\begin{algorithm}[H]
\caption{Construct Goal History}
\label{alg_goalHistory}
\footnotesize
\begin{algorithmic}[1]
\Procedure {ConstructHistory}{Goal $ G = (G_{id}, N_{id}) $}
	\ForEach {$ t_i $}
		\State log time instance $ t_i $
		\State log ObjectiveFunction(client $ c $)
		\State log executed Tactic $ T_x $
		\State log ExecutionTrace $ \tau(G) $
		\State get GoalSatisfaction $ v(G) $ 
		\If {InteractionAwareness is enabled}
			\State log Environment Goals $ G_e = \{ G_1(n_1, i), G_2( n_2, i), ..., G_x(n_x, i) \} $
			\State log Environment Tactics $ T_e = \{ T_1, T_2, ..., T_x \} \mbox{ for } \forall \mbox{ } G \in G_e $
		\EndIf	
	\EndFor
\EndProcedure
\end{algorithmic}
\end{algorithm}
\end{minipage}
\par
\end{center}

\subsection{Time-Awareness for Stability Online Learning}
\label{sec_selfAwareness_time}
Learning from historical information about adaptation actions and stability states, time-awareness is capable to take adaptation decisions converging towards stability. We use a form of model-free reinforcement learning technique, that is ``Q-learning''. The technique does not require a model of the system (i.e. priori knowledge and computational demands) and can handle problems with stochastic transitions with the capability of learning how to act optimally in a controlled Markovian context \cite{Watkins1992} \cite{Dayan2006} \cite{Poole2010}. 

Given the runtime uncertainty, we consider the system as a finite Markov decision process (FMDP), where the Q-learning can identify an optimal action-selection policy (i.e. adaptation action), where the expected value of the total reward return is the maximum achievable at the current state. The technique works during runtime by successively improving its evaluations of the quality of particular adaptation actions at particular states (i.e. online learning) \cite{Watkins1992} \cite{Dayan2006} \cite{Poole2010}. 

\subsubsection{Learning Model}
Formally, the learning process involves a set of states $ S $ and a set of adaptation actions $ A $, where each state $ s \in S $ present the status of stability attributes and each action $ a \in A $ is one of the possible different configurations of the architecture. A state $ s $
is a tuple of the different stability attributes, as such $ <rt, c, ...> $ for response time and cost are the stability attributes. An action $ a $ is a tuple of the architecture configuration settings subject to adaptation, such as number of PMs and number of VMs $ <pm, vm, ...> $.

When performing an action $ a \in A $, the system transitions from state $ s_t $ to state $ s_{t+1} $. Executing an action in a specific state is evaluated by a reward $ r $ value. The Q-Learning algorithm has a function that calculates the quality of a state-action combination:
\begin{equation}
Q : S \times A \rightarrow \mathbb{R}
\end{equation}

where $ \mathbb{R} $ is the reward Q-matrix. The matrix is in the following format:
\[
\mathbb{R} = 
\begin{bmatrix}
  r_{s_1,a_1} & r_{s_1,a_2} &  \dots & r_{s_1,a_n} & \\
  r_{s_2,a_1} & r_{s_2,a_2} & \dots & r_{s_2,a_n} & \\
  & \ddots & & & \\
  r_{s_m,a_1} & r_{s_m,a_2} & \dots & r_{s_m,a_n} & \\
\end{bmatrix}
\]

where the rows of the matrix represent the different states $ s $ of the system, the columns represent the possible adaptation actions $ a $, and the matrix values are the learnt reward values $ r $. 

The $ Q $ function returns the reward used to provide the reinforcement and stand for the quality of an action taken in a given state. At each time instance $ t $, the algorithm selects an action $ a_t $, observes a reward $ r_t $, enters to a new state $ s_{t+1} $ that depends on the previous state $ s_t $ and the selected action $ a_t $, and the Q-matrix is updated using the weighted average of the old value and the new information, as follows:
\begin{equation} \label{eq_qlearning}
Q(s_t, a_t) \leftarrow (1 - \alpha) \: Q(s_t, a_t) + \alpha \: \Big(r_t + \gamma \; \max_{a}Q(s_{t+1},a)\Big)
\end{equation}

where $ r_t $ is the reward observed for the current state $ s_t $, $ \alpha $ is the constant learning rate $ (0 < \alpha \leq 1) $ that determines the extent to which the newly acquired information overrides old information, and $ \gamma $ is the discount factor $ (0 < \alpha \leq 1) $ that determines the importance of future rewards. If the learning factor $ \alpha $ is set $ = 1 $, the algorithm uses only the most recent information, and if $ \alpha = 0 $, this forces the algorithm to learn nothing and use historical information only. If the discount factor $ \gamma = 0 $, the algorithm will consider only current rewards, and if $ \gamma = 1 $, the algorithm will use long-term high reward, which is the case of stability.

\subsubsection{Online Learning Algorithm}
The goal of the learning algorithm is to maximise the total reward in the future, by learning which action is optimal for each state. The optimal action for each state is the one that has the highest long-term reward, in order to achieve stability on the long-run. The reward is a weighted sum of the expected values of the rewards of all future steps starting from the current state.

The matrix is initialised with a possibly arbitrary fixed value. For simplicity, we assume a certain number of states with ranges of each stability attribute and a certain set of configurations. These ranges of stability attributes and possible configurations could be easily refined adding more columns and rows in the Q-matrix.

The algorithm keeps running while the system is online. First, the current state is observed, and the algorithm selects an action with the highest reward value among the set of actions for the current state using equation \ref{eq_qlearning}. Then, the new state and new reward are observed, and the Q-matrix is updated with new reward value.

\begin{center}
\begin{minipage}{0.85\linewidth}
\footnotesize
\begin{algorithm}[H]
\renewcommand{\algorithmicrequire}{\textbf{Input:}}
\renewcommand{\algorithmicensure}{\textbf{Output:}}
\algnewcommand{\Local}{\BState\textbf{Local variables: }}
\caption{Q-Learning}
\label{alg_runtimeGoalInstance}
\begin{algorithmic}[1]
\Procedure {QLearning}{$ S, A, \alpha, \gamma $} \\
	\algorithmicrequire		\\
		\hspace*{\algorithmicindent} $ S : $  set of states	\\
		\hspace*{\algorithmicindent} $ A : $  set of actions 	\\
		\hspace*{\algorithmicindent} $ \gamma : $  discount factor 	\\
		\hspace*{\algorithmicindent} $ \alpha : $  learning rate	\\
	\algorithmicensure		\\
		\hspace*{\algorithmicindent} $ a' : $  new action	
	\Local		\\
		\hspace*{\algorithmicindent} $ Q[S, A] : $  Q-matrix	\\
		\hspace*{\algorithmicindent} $ s : $  previous state	\\
		\hspace*{\algorithmicindent} $ a : $  previous action	\\
		\hspace*{\algorithmicindent} $ r : $  reward	\\	
		\hspace*{\algorithmicindent} $ s' : $  new state	
\BState \emph{initialise}:		
	\State $ S = \{s_1<rt, c, ...>, s_2<rt, c, ...>, ... \} $
	\State $ A = \{a_1<pm, vm, ...>, a_2<pm, vm, ...>, .... \} $
	\State get current state $ s $
	\Repeat	\Comment{while online}	
		\State select action $ a' $ from possible actions for $ s $
		\State $ Q(s, a) \leftarrow (1 - \alpha) \: Q(s, a) + \alpha \: (r + \gamma \; max_{a'}Q(s',a')) $
		\State observe reward $ r $ and new state $ s' $	
		\State $ s \leftarrow s' $
	\Until{termination}
\EndProcedure
\end{algorithmic}
\end{algorithm}
\end{minipage}
\par
\end{center}

\subsection{Meta-Self-Awareness for Managing Trade-offs between Stability Attributes}
\label{sec_selfAwareness_meta}
We investigate the use of game theory to achieve an equilibrium point between different stability quality attributes, i.e. modelling and analysing the consequent trade-offs between stability attributes given the uncertainty of the running environment. The proposed methodology considers the value implications of choosing an architectural tactic for adaptation with respect to multiple quality attributes subject to stability and potentially uncertain future runtime conditions. In more details, we tend to evaluate architectural tactics for their pay-off values and based on such an evaluation, an architectural tactic is selected in a way that supports the management of trade-offs between different stability goals. The goal is to select the tactic for better adaptation leading to the long-term welfare of the architecture. Architectural tactics are intended to aid in creating architectures that meet quality requirements \cite{Procaccianti2014}. Such tactics are employed to achieve a desired quality attribute behaviour, which, in turn, imparts utility to the architecture. The utility should not be in terms of one quality attribute, yet an aggregate utility comprehending multiple quality attributes.

We consider the continuous runtime process of managing trade-offs under the uncertainty of the environment as a \textit{stochastic game}, where the players are the runtime stability attributes and their strategies are the possible adaptation actions. A central idea is that architectural decision, such as the application of a tactic for adaptation, is analogous to a game strategy. Quality attributes and their expected utility under uncertainty act as underlying assets for the valuation of architectural decisions, similar to the valuation of game strategies. This approach provides quantitative decision for selecting architectural tactic based on the utility objectives and uncertainty of runtime workloads, quality goals and environmental changes. Part of the objectives is to evaluate the overall adaptation process and its implication for the long-term welfare of the architecture and goals fulfilment.

\subsubsection{Problem formulation}
Achieving runtime architectural stability among different stability attributes should involve a careful understanding of the relationship, impact, correlation and sensitivity among attributes subject to stability, as well as handling potential conflicts. Given the runtime uncertainty arising from many sources, the runtime stability is seen to be a probable behaviour rather than deterministic.

The proposed approach builds upon the framework for modelling, analysing and automatic verification of turn-based stochastic multi-players games (SMGs) \cite{Chen2013b}. A natural fit for modelling systems that exhibit probabilistic behaviour is using stochastic games \cite{Chen2013b} \cite{Chen2013a}. Probabilistic model checking provides verification of quantitative properties (stability gaols) and provides a means to synthesise optimal strategies to achieve these goals \cite{Chen2013a} and leave the architecture stable on the long-run. 

A natural fit for modelling systems that exhibit probabilistic behaviour is adopting a game-theoretic perspective \cite{Chen2013b} \cite{Chen2013a}. In particular, \textit{stochastic games} can used to model the self-adaptive (stochastic) system and its (conflicting) stability goals. Probabilistic model checking provides a means to model and analyse these systems, by providing verification of quantitative properties in probabilistic temporal logic \cite{Chen2013a}. PRISM-games tool, built on the code-base of PRISM model checker, provides modelling of quantitative verification for SMGs, where the games are specified using the PRISM modelling language \cite{Chen2013a}. In this tool, SMGs are described as a model composed of \textit{modules}, where their state is determined by a set of \textit{variables} and their behaviour is specified by a set of \textit{guarded commands}, containing an optional action label, a guard and a probabilistic update for the module variables \cite{Chen2013a}:
\[ [\mathtt{action}] \; \; \mathtt{guard} \; \; -> \; \; \mathtt{prob_1} \; : \; \mathtt{update_1} \; \; + \; \; \dots \; \; + \; \; \mathtt{prob_n} \; : \; \mathtt{update_n}  \]

PRISM-games' properties specification are written using a probabilistic temporal logic with rewards called rPATL \cite{Chen2013a} \cite{Chen2013b}. rPATL is an extension of the logic PATL \cite{Chen2007}, which is itself an extension of ATL \cite{Alur2002}, a widely used logic for reasoning about multi-player games and multi-agent systems \cite{Chen2013a}. Properties, quantitatively specified, in rPATL can state that a coalition of players has a strategy which can ensure that either the probability of an event's occurrence or an expected reward measure meets some threshold \cite{Chen2013b}. rPATL is a CTL-style branching time temporal logic that incorporates the coalition operator $ \langle\langle C \rangle \rangle $ of ATL \cite{Alur2002}, the probabilistic operator $ \mathsf{P}_{\bowtie q} $ of PCTL \cite{Bianco1995}, and the reward operator $ \mathsf{R}_{\bowtie x}^{r} $ from \cite{Forejt2011} for reasoning goals related to reward/cost measures \cite{Chen2013a}. Beside the \textit{precise} value operators, rPATL also supports the quantification of maximum and minimum accumulated reward until a $ \phi $-state is reached that can be guaranteed by players in coalition $ C $, noted as $ \langle \langle {C} \rangle \rangle \mathsf{R}^{r}_{max=?} [\mathsf{F}^{*}\phi] $ and $ \langle \langle {C} \rangle \rangle \mathsf{R}^{r}_{min=?} [\mathsf{F}^{*}\phi] $ respectively.

By expressing properties that enable us to quantify the maximum and minimum rewards a player can achieve, we can reason about different adaptation strategies and synthesise strategies that optimise stability rewards. This allows to choose an optimal adaptation action that would achieve stability attributes, and hence leave the architecture in a stable state on the long-run. The approach consists of SMG model (discussed in section \ref{sec_selfAwareness_meta_gameModel}) and strategy synthesis (section \ref{sec_selfAwareness_meta_strategy}). Then, we describe the model specification (in section \ref{sec_selfAwareness_meta_modelSpecs}).

\subsubsection{Stochastic Multi-Player Game Model}
\label{sec_selfAwareness_meta_gameModel}
We model the self-adaptive system and its environment as two players of a SMG, in which the system's objective is reaching stability state, that is a goal state that maximises a utility/reward (i.e. achieve stability attributes), and the environment as an opponent whose actions cannot be controlled. In each turn, only one player can choose between different strategies, and the outcome can be probabilistic. The system can choose between a set of adaptation actions, i.e. adaptation tactics, to achieve stability gaols, while the environment is considered as an adversary to the system.

\theoremstyle{definition}
\begin{definition}{(SMG)}
A \textit{turn-based stochastic multi-player game} (SMG) is a tuple $ \mathcal{G} = \langle \Pi, \mathnormal{S}, \mathnormal{A}, (\mathnormal{S_i})_{i \in \pi}, \Delta, \mathnormal{AP}, \mathcal{X}, \mathnormal{r} \rangle $, where $ \Pi $ is a finite set of players, $ \mathnormal{S} \neq \emptyset $ is a finite set of states, $ \mathnormal{A} \neq \emptyset $ is a finite set of actions, $ (\mathnormal{S_i})_{i \in \pi} $ is a partition of $ \mathnormal{S} $, $ \Delta : \mathnormal{S} \times \mathnormal{A} \rightarrow \mathcal{D}(\mathnormal{S}) $ is a partial transition function ($ \mathcal{D}(\mathnormal{S}) $ denotes the set of discrete probability distributions over finite set $ \mathnormal{S} $), $ \mathnormal{AP} $ is a finite set of atomic propositions, $ \mathcal{X} : \mathnormal{S} \rightarrow 2^{\mathnormal{AP}} $ is a labeling function, and $ \mathnormal{r} : \mathnormal{S} \rightarrow \mathbb{Q}_{\geq 0} $ is a reward structure mapping each state to a non-negative rational reward.
\end{definition}

In each state $ s \in \mathnormal{S} $, the set of available actions is denoted by $ \mathnormal{A}(s) = \{ a \in \mathnormal{A} | \Delta(s,a) \neq \bot \} $, assuming that $ \mathnormal{A}(s) \neq \emptyset $ for all states. The choice of action in each state $ s $ is under control of one player $ i \in \Pi $, for which $ s \in \mathnormal{S}_i $. 

The set of players $ \Pi = \{ sys, env \}$ is formed by the self-adaptive system and its environment. The set of states $ \mathnormal{S} = \mathnormal{S}_{sys} \cup \mathnormal{S}_{env} $ is formed of the states of the system $ \mathnormal{S}_{sys} $ and the states of the environment $ \mathnormal{S}_{env} $ ($ \mathnormal{S}_{sys} \cap \mathnormal{S}_{env} \neq \emptyset $). The set of actions $ \mathnormal{A} = \mathnormal{A}_{sys} \cup \mathnormal{A}_{env} $ is formed of the set of actions available for the system and the environment denoted by $ \mathnormal{A}_{sys} $ and $ \mathnormal{A}_{env} $ respectively. $ \mathnormal{AP} $ is the subset of all predicates that can be built over the state variables and includes the \textit{goal} that is satisfied when achieving stability goals. 

\theoremstyle{definition}
\begin{definition}{(Path)} A \textit{path} of SMG $ \mathcal{G} $ is a possibly infinite sequence $ \lambda = s_0a_0s_1a_1 \dots $, such that $ a_j \in \mathnormal{A}(\mathnormal{s}_j) $ and $ \Delta (\mathnormal{s}_j,\mathnormal{a}_j)(\mathnormal{s}_{j+1}) > 0 $ for all $ j $. $ \Omega^+_ \mathcal{G} $ is used to denote the set of finite states in $ \mathcal{G} $.
\end{definition} 

$ r $ denotes the reward for labelling goal states with their associated utility. The reward of a state $ s $ is defined as $ r(s) = \Sigma^q_{i=1}u_i(v_i^s) $ if $ s \models $ (satisfies) \textit{goal}, where $ u_i \in [0,1] $  is the utility function for the stability goal $ i \in \{1,\dots,q\} $, and $ v_i^s $ is the value of the state variable associated with the architectural property representing stability attribute $ i $ in state $ s $.

\theoremstyle{definition}
\begin{definition}{(Reward structure)} A reward structure for $ \mathcal{G} $ is a function $ r : S \rightarrow \mathbb{R}_{\geq 0} $ or $ r : S \rightarrow \mathbb{R}_{\leq 0} $.
\end{definition}

The reward structure is used to maximise or minimise the goals. A reward structure assigns values to pairs of states and actions. 

Players of the game can follow strategies for choosing actions that result in achieving their goals.

\theoremstyle{definition}
\begin{definition}{(Strategy)} A \textit{strategy} for player $ i \in \Pi $ in $ \mathcal{G} $ is a function $ \sigma_i : (\mathnormal{SA})*\mathnormal{S}_i \rightarrow \mathcal{D}(\mathnormal{A}) $ which, for each path $ \lambda . s \in \Omega^+_ \mathcal{G} $ where $ s \in \mathnormal{S}_i $, selects a probability distribution $ \sigma_i(\lambda . s) $ over $ \mathnormal{A}(s) $.
\end{definition}

A strategy $ \sigma_i $ is memoryless if $ \sigma_i(\lambda.s) = \sigma_i(\lambda'.s) $ for all paths $ \lambda.s, \lambda'.s \in \Omega^+_ \mathcal{G} $, and deterministic if $ \sigma_i(\lambda.s) $ is a Dirac distribution for all $ \lambda.s \in \Omega^+_ \mathcal{G} $.

\subsubsection{Strategy Synthesis}
\label{sec_selfAwareness_meta_strategy}
Reasoning about \textit{strategies} is an fundamental aspect of SMGs model checking. rPATL queries check for the existence of a strategy that is able to optimise an objective or satisfies a given probability/reward bound \cite{Chen2013a}. Model checking also supports optimal \textit{strategy synthesis} \cite{Chen2013a} for a given property. In our case, we use \textit{memoryless deterministic strategies}, that resolve the choices in each state selecting actions based on the current state \cite{Chen2013a}. Such strategies are guaranteed to achieve the optimal expected rewards \cite{Chen2013a}. 

We perform strategy synthesis using \textit{multi-objectives queries} supported by PRISM-games 2.0, by computing Pareto set or optimal strategies for managing trade-offs between multi-objective properties \cite{Kwiatkowska2016}. Multi-objectives queries are expressed as boolean combination of reward-based objectives with appropriate weights \cite{Kwiatkowska2016}, which allows reasoning about long-run average reward. Generally, higher weights are given to the stability of quality of service attributes (e.g. response time), as these are the main objective of adaptation. 

Properties are specified as follows:
$ \langle \langle {sys} \rangle \rangle \mathsf{R}^{r}_{max=?} [\mathsf{F}^{c}\phi] $, to synthesise a strategy that maximises the utility rewards from all stability attributes, where $ \phi $ state represents the state where adaptation goals are achieved. The multi-objective query to reason about stability multi-objective property is specified as follows:
\[ \langle \langle {sys} \rangle \rangle (\mathsf{R}\{ response\_time \}_{\leq v_1}[\mathsf{C}] \land \mathsf{R}\{ energy \}_{\leq v_2}[\mathsf{C}]) \]

where the targets $ v_1, v_2, \dots $ for the stability objectives are defined from Service Level Agreements (SLAs).

\subsubsection{Model Specification}
\label{sec_selfAwareness_meta_modelSpecs}
Our formal model is implemented using PRISM-games 2.0 \cite{Chen2013a} \cite{Kwiatkowska2016}. The state space and behaviours of the game are generated from the stochastic processes under the control of the two players of the game, the system and the environment. In more details:

\paragraph{The self-adaptive system (player \textit{sys})}
controls the process that models the adaptation controller of the self-adaptive system, which is responsible about triggering and executing adaptation actions. The set of actions available to the system $ A_{sys} $ are the set of adaptation tactics defined in the adaptation controller, e.g. horizontal scaling, vertical scaling, increasing VM capacity. Each action $ a \in A_{sys} $ command follows the pattern:
\begin{multline}
[\mathtt{a}] \; \; \mathtt{C_a} \wedge \neg \mathtt{goal} \wedge \mathtt{t} = \mathnormal{sys} \; \; -> \; \; \mathtt{prob_a^1} \; : \; 
\\
\mathtt{update_a^1}\wedge \mathtt{t'} = \mathnormal{env} \; \; + \; \; \dots \; \; + \; \; \mathtt{prob_a^n} \; : \; \mathtt{update_a^n}\wedge \mathtt{t'} = \mathnormal{env}  
\end{multline}

where $ \mathtt{C_a}  $ is the constraints for executing the tactic $ \mathtt{a} $ (e.g. capacity of a physical machine (PM) to accommodate virtual machines (VMs)), a predicate $ \neg \mathtt{goal} $ to prevent expanding the state space beyond the satisfaction of the adaptation goal, $ \mathtt{t} = \mathnormal{env} $ constraints the execution of actions of the player in turn $ \mathtt{t} $ to states $ s \in S_{sys s} $. The command includes the possible updates $ \mathtt{update_a^i} $, corresponding to one probabilistic outcome for the execution of $ \mathtt{a} $, along with their associated probabilities $ \mathtt{prob_a^i} $. And the turn is given back to the $ \mathnormal{env} $ player by the control variable $ \mathtt{t'} $.

\paragraph{The environment (player \textit{env})} 
controls the process that models potential disturbances to the stability of the system that are out of the system's control, e.g. VM failure, server fault, network latency. The environment process is specified as a set of commands with asynchronous actions $ a \in A_{env} $, and its local choices are specified non-deterministically to obtain a rich specification of the environment's behaviour. Each command follows the pattern:
\begin{multline}
[\mathtt{a}] \; \; \mathtt{C_a^e} \wedge \neg \mathtt{end} \wedge \mathtt{t} = \mathnormal{env} \; \; -> \; \; \mathtt{prob_a^1} \; : \;
\\
 \mathtt{update_a^1}\wedge \mathtt{t'} = \mathnormal{sys} \; \; + \; \; \dots \; \; + \; \; \mathtt{prob_a^n} \; : \; \mathtt{update_a^n}\wedge \mathtt{t'} = \mathnormal{sys}  
\end{multline}

where $ \mathtt{C_a^e}  $ is the environment constraints for the execution of action $ \mathtt{a} $,  $ \neg \mathtt{end} $ prevents the generation of further states, and $ \mathtt{t} = \mathnormal{env} $ constraints the execution of actions of the player in turn to states $ s \in S_{env} $. The command includes the possible updates $ \mathtt{update_a^i} $, corresponding to one probabilistic outcome for the execution of $ \mathtt{a} $, along with their associated probabilities $ \mathtt{prob_a^i} $. And the turn is given back to the system player.

The SMG model consists of the following modules: 

\paragraph{Players definition.}
Listing \ref{code_players} shows the definition of the stochastic game players: player $ \mathsf{env} $ which is control of the actions that the system environment can take, and player $ \mathsf{sys} $ which controls the actions to be taken by the adaptation controller and the execution of adaptation tactics. The global variable $ \mathsf{t} $ is used to control turns in the game, alternating between the system and the environment.

\begin{lstlisting}[frame=tb, caption={Players definition in PRISM-games 2.0}, label=code_players]
player env environment [] endplayer
player sys system [], [] endplayer
const TURN_SYS, TURN_ENV;
global t:[TURN_SYS..TURN_ENV] init TURN_ENV;
\end{lstlisting}

\paragraph{Environment.}
The \textsf{environment} module (encoding shown in Listing \ref{code_environment}) allows to obtain a representative specification of the system's environment, introducing disturbance to the stability of the system. This is done using variables that represent configurations that might affect stability, e.g. changing number of VMs, changing number of PMs. These behaviours are parametrised by the constants: \textsf{MAX\textunderscore TOTAL\textunderscore VM\textunderscore NUM} and \textsf{MAX\textunderscore TOTAL\textunderscore PM\textunderscore NUM} that constraints the maximum number of VMs and PMs respectively that the environment can use to introduce disturbance, \textsf{MAX\textunderscore TOTAL\textunderscore VM\textunderscore CAP} and \textsf{MAX\textunderscore TOTAL\textunderscore PM\textunderscore CAP} that constraints the maximum capacity of VMs and PMs respectively, \textsf{MAX\textunderscore VM\textunderscore CHANGE} is the maximum numbers of virtual machines (VMs) that the environment can change to interrupt the system execution and cause instability, \textsf{MAX\textunderscore PM\textunderscore CHANGE} is the maximum number of physical machines (PMs) that the environment can change to cause instability in QoS provision (e.g. response time). For simplicity, we consider all PMs and VMs are of the same capacity.

The current state of the environment is defined using the variables: \textsf{current\textunderscore vm\textunderscore num}, \textsf{current\textunderscore pm\textunderscore num} corresponding to the changes introduced by the environment at the current turn with respect to the number of VM and PM respectively, \textsf{total\textunderscore vm\textunderscore cap} and \textsf{total\textunderscore pm\textunderscore cap} that keep track of the total capacity of VM and PM respectively.

At each turn, the environment action is setting the disturbance variables (changing system configurations) using the command in Listing \ref{code_environment} line 12. First, the guard checks that: (i) it is the turn of the environment (\textsf{t=TURN\textunderscore ENV}), (ii) an absorbing state has not been reached yet (\textsf{!end}), and (iii) the total number of VMs and PMs as well as their total capacities will not exceed the maximum specified for all types of disturbance. If the guard conditions are satisfied, the command: (i) sets the current configuration variables (e.g. $d_{vm}$), (ii) updates the total capacity variables with the current disturbance variables, and (iii) gives the turn to the system (\textsf{t'=TURN\textunderscore SYS}).

\begin{lstlisting}[frame=tb, caption={Environment module}, label=code_environment]
const MAX_VM_CHANGE, MAX_PM_CHANGE;
const MAX_TOTAL_VM_NUM, MAX_TOTAL_PM_NUM, 
		MAX_TOTAL_VM_CAP, MAX_TOTAL_PM_CAP;

module environment
current_vm_num: [1..MAX_TOTAL_VM_NUM] init 1;
current_pm_num: [1..MAX_TOTAL_PM_NUM] init 1;
total_vm_cap: [1..MAX_TOTAL_VM_CAP] init 1;
total_pm_cap: [1..MAX_TOTAL_PM_CAP] init 1;
[] (t=TURN_ENV)&(!end)& 
	($d_{vm}$<MAX_VM_CHANGE)&
	($d_{pm}$<MAX_PM_CHANGE)&
	($d_{vm}$+current_vm_num<MAX_TOTAL_VM_NUM)& 
	($d_{pm}$+current_pm_num<MAX_TOTAL_PM_NUM) -> 	
	(current_vm_num=current_vm_num+$d_{vm}$)& 
	(current_pm_num=current_pm_num+$d_{pm}$)& 
	(total_vm_cap=current_vm_num*cap)&
	(total_pm_cap=current_pm_num*cap)&
	(t'=TURN_SYS);
endmodule
\end{lstlisting}

\paragraph{System.}
The \textsf{system} module models the behaviour of the system, including the adaptation controller and the execution of adaptation tactics (Listing \ref{code_system}). This is parametrised by the constants: (i) \textsf{MIN\textunderscore PM\textunderscore NUM} and \textsf{MAX\textunderscore PM\textunderscore NUM} whcich specify the minimum and maximum number of PMs, (ii) \textsf{MIN\textunderscore VM\textunderscore NUM} and \textsf{MAX\textunderscore VM\textunderscore NUM} which are the minimum and maximum number of VMs that PMs can accommodate, (iii) \textsf{MIN\textunderscore PM\textunderscore CAP} and \textsf{MAX\textunderscore PM\textunderscore CAP} is the minimum and maximum computational capacity of a PM configuration, (iv) \textsf{MIN\textunderscore VM\textunderscore CAP} and \textsf{MAX\textunderscore VM\textunderscore CAP} is the minimum and maximum computational capacity of a VM configuration, (v) \textsf{STEP\textunderscore NUM} and \textsf{STEP\textunderscore CAP} which are used to increase or decrease configuration, and (vi) \textsf{INIT\textunderscore PM\textunderscore NUM}, \textsf{INIT\textunderscore VM\textunderscore NUM}, \textsf{INIT\textunderscore VM\textunderscore CAP}, \textsf{INIT\textunderscore PM\textunderscore CAP} for the initial configuration of the architecture with respect to PMs, VMs and VMs capacity.

The variables of the module represent the current configuration of the architecture (\textsf{pm\textunderscore num}, \textsf{vm\textunderscore num}, \textsf{pm\textunderscore cap}, \textsf{vm\textunderscore cap}), the current provisioned quality of service (\textsf{respnse\textunderscore tine}, \textsf{energy}, \textsf{cost}), and quality of adaptation (\textsf{settling\textunderscore time}, \textsf{resources\textunderscore overshoot}, \textsf{adaptation\textunderscore frequency}). To update the value of quality variables, we employ multiple $ M/M/1 $ queueing model (from our earlier work \cite{Salama2018b}) to compute them based on the current architecture configuration and the request arrivals. 

\begin{lstlisting}[frame=tb, caption={System module}, label=code_system]
const MIN_PM_NUM, MAX_PM_NUM, MIN_VM_NUM, MAX_VM_NUM, 
	MIN_VM_CAP, MAX_VM_CAP,  MIN_PM_CAP, MAX_PM_CAP, 
	STEP_NUM, STEP_CAP;
const INIT_PM_NUM, INIT_VM_NUM, INIT_PM_CAP, INIT_VM_CAP;

module system
pm_num: [1..MAX_PM_NUM] init INIT_PM_NUM;
vm_num: [1..MAX_VM_NUM] init INIT_VM_NUM;
pm_cap: [1..MAX_PM_CAP] init INIT_PM_CAP;
vm_cap: [1..MAX_VM_CAP] init INIT_VM_CAP;

respnse_tine, energy, cost;
settling_time, resources_overshoot, adaptation_frequency;

[] (t=TURN_SYS)&(goal)&(!end) -> (t'=TURN_ENV);
	[increase_pm_num](pm_num<MAX_PM_NUM) -> 
		(pm_num=pm_num+STEP_NUM);
	[decrease_pm_num](pm_num>MIN_PM_NUM) -> 
		(pm_num=pm_num-STEP_NUM);
	[increase_vm_num](vm_num<MAX_VM_NUM) -> 
		(vm_num=vm_num+STEP_NUM);
	[decrease_vm_num](vm_num>MIN_VM_NUM) -> 
		(vm_num=vm_num-STEP_NUM);
	[increase_vm_cap](vm_cap<MAX_VM_CAP) -> 
		(vm_cap=vm_cap+STEP_CAP);
	[decrease_vm_cap](vm_cap>MIN_VM_CAP) -> 
		(vm_cap=vm_cap-STEP_NUM);
endmodule
\end{lstlisting}

\paragraph{Properties and Rewards.}
To perform adaptations leading to stability and managing trade-offs between its attributes, we use rPATL for the specification stability properties. These properties are used as input to PRISM- games, which can synthesise optimal adaptation actions for the attributes subject to stability. We use \textit{long-run} properties from PRISM-games 2.0 (an extension for PRISM-games) \cite{Kwiatkowska2016}, which allow expressing properties of autonomous systems that run for long periods of time and specify measures, such as energy consumption per time unit \cite{Kwiatkowska2016}. 

The effect of adaptation strategies on stability goals is encoded using a reward structure that assigns real-values of stability goals \cite{Kwiatkowska2016}. We use long-run average reward for expressing cumulative rewards towards stability. Each stability goal has a \textit{target} value $ \mathtt{v} $ for a reward value as a maximum or minimum. Goals for the expected long-run average reward $ \mathtt{r} $ is expressed as $ \mathtt{R\{``r"\}_{\geq v}[S]} $, where $ \mathtt{S} $ denotes long-run rewards. Satisfaction objectives for long-run rewards are expressed as $ \mathtt{P_{\geq 1} [R(path) \{``r"\}_{\geq v} [S]] } $.

\section{Experimental Evaluation}
\label{sec_evaluation}
The main objective of the experimental evaluation is to examine stability when using different self-awareness capabilities for reasoning about stability, and assess associated overhead. First, we briefly introduce the architecture's domain and experiments setup, then experimentally evaluate stability attributes and adaptation properties. 

\subsection{Architecture Domain}
\label{sec_evaluation_domain}
Cloud-based software architectures are a suitable example of dynamism, unpredictability and uncertainty \cite{Armbrust2010}. The execution environment of cloud architectures is highly dynamic, due to the on-demand nature of the cloud. Cloud architectures operate under continuous changing conditions, e.g. changes in workload (number/size of requests), end-user quality requirements, unexpected circumstances of execution (peak demand) \cite{Chen2014a} \cite{Villegas2017}. The on-demand service provision in clouds imposes performance unpredictability and makes the elasticity of resources an operational requirement. 

Due to the on-demand and dynamic nature of cloud, there is an increasing demand on cloud services, where the realisation of quality requirements should be managed without human interventions. This type of architecture tends to highly leverage on adaptation (e.g. changing behaviour, reconfiguration, provisioning additional resources, redeployment) to regulate the satisfaction of end-users' requirements under the changing contexts of execution \cite{Salehie2009} \cite{Villegas2017}. The self-adaptation process is meant to make the system behaviour converges towards the intended behaviour, i.e. quality requirements of the end-users without SLA violation \cite{Villegas2017}. The purpose of adaptation is to satisfy the runtime demand of multi-tenant users, by changing configuration and choosing optimal tactics for adaptation. An unstable architecture will risk not improving or even degrading the system to unacceptable states \cite{Villegas2011}. In such case, there are more dynamics to observe, and stability is challenging with the continuous runtime adaptations in response to the perception of the execution environment and the system itself \cite{Villegas2017}. 

Further, the economic model of clouds (pay-as-you-go) imposes on providers economic challenges for SLA profit maximisation by reducing their operational costs \cite{Armbrust2010}. Also, providers face monetary penalties in case of SLA violations affecting their profit, which push them towards stabilising the quality of service provisioned. With the rising demand of energy, increasing use of IT systems and potentially negative effects on the environment, the environmental aspect (in terms of energy consumption) has emerged as a factor affecting the software quality and sustainability \cite{Lago2015}. While sometimes imposed by laws and regulations, decreasing energy consumption does not have only potential financial savings, but also affects the ecological environment and the human welfare \cite{Lago2015}. So, environmental requirements should be considered and traded off against business requirements and financial constraints \cite{Lago2015}. 

\subsection{Experiments Setup}
\label{sec_evaluation_setup}
\begin{figure}[!h]
\centering
\includegraphics[width=\textwidth]{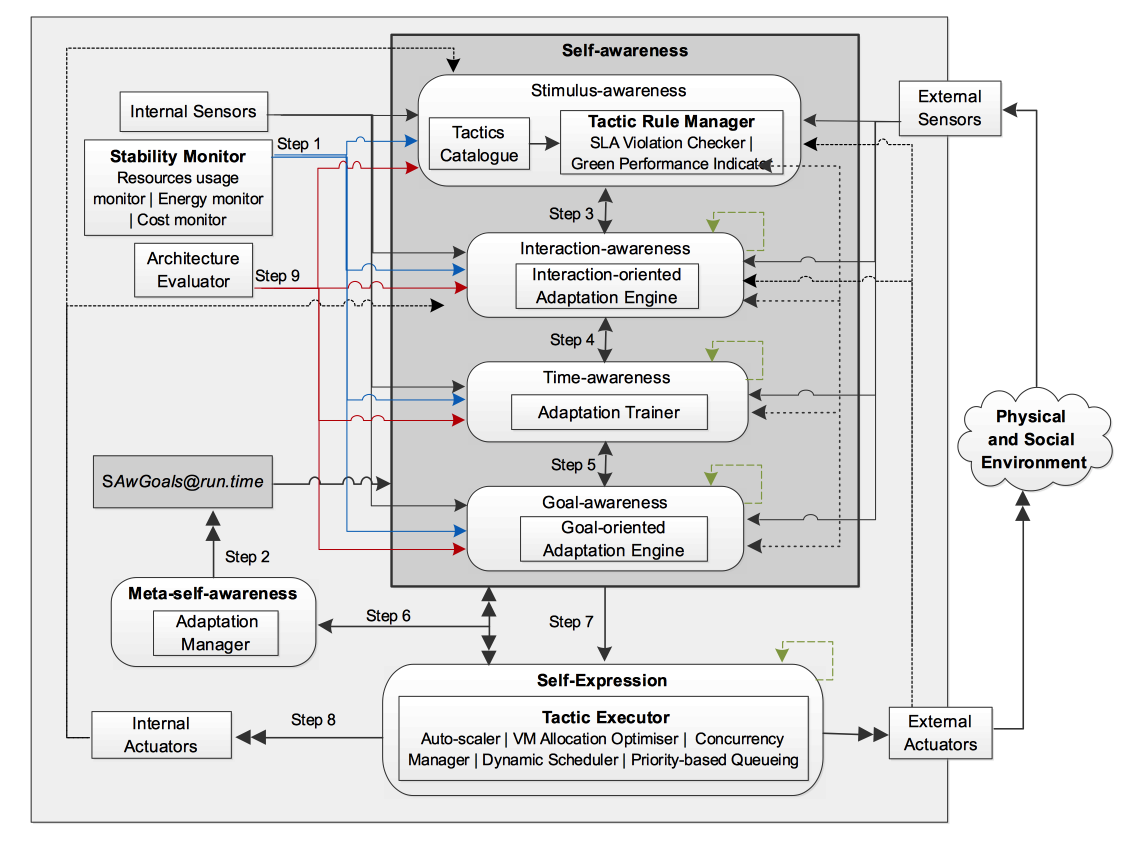}
\caption{Architecture of the instantiated Cloud node}
\label{fig_cloudArchitecture}
\end{figure}

To conduct the experimental evaluation, we implemented the architecture of a cloud node using the widely adopted \textit{CloudSim} simulation platform for cloud environments \cite{Calheiros2011}. The simulation was built using Java JDK 1.8 and was run on a 2.9 GHz Intel Core i5 16 GB RAM computer. The architecture of the cloud node is illustrated in Figure \ref{fig_cloudArchitecture}. 

The architecture embeds Stability Monitor component for monitoring stability. Components necessary for checking possible violation of stability attributes are implemented in the stimulus-awareness component. Tactics are defined in the Tactics Catalogue component in the stimulus-awareness component. Management components of tactics were configured into the Tactic Executor for running the tactics, e.g. auto-scaler. The proposed self-awareness techniques are implemented in the corresponding components, i.e. SAw-Goals@run.time, goals management, online learning. Regarding the trade-offs management, we implemented the case in \textsf{PRISM-Games 2.0.beta3} running on the same machine with OS X10.13.4. Then, the outcome strategies were exported to the simulator for performing adaptations.

The stability attributes, to be taken into consideration in this case (as defined in \cite{Chen2014a}, \cite{Salama2017b}), include: (i) quality requirements specified in end-users SLAs, (ii) environmental restrictions, (iii) economic constraints, and (iv) quality of adaptation. Table \ref{tbl_StabilityAttributes} lists details of the stability attributes. With respect to the quality requirements, we consider performance (measured by response time from the time the user submits the request till the cloud submits the response back to the user in milliseconds). For the environmental aspect, we use the greenability property \cite{Lago2015} \cite{Calero2015} measured by the amount of energy consumed for running hosts in kWh. For the economic constraints, we define the operational cost by the cost of computational resources (CPUs, memory, storage and bandwidth for running VMs). Regarding adaptation properties, we consider the frequency of adaptation (number of adaptation cycles performed) and adaptation overhead (time consumed in the adaptation process) and resources overshoot (computational resources used in the adaptation process) \cite{Villegas2011} \cite{Villegas2017}.

\begin{table}[!h]
\caption{Stability attributes}
\label{tbl_StabilityAttributes}
\center
\footnotesize
\begin{tabularx}{0.75\textwidth}
{
>{\raggedright}p{0.27\textwidth} 
>{\raggedright}p{0.10\textwidth} 
>{\raggedright}p{0.15\textwidth} 
>{\raggedright\arraybackslash}p{0.10\textwidth} 
}
\toprule
	\textbf{Attribute}						&
	\textbf{Weight}							&
	\textbf{Metric}								&	
	\textbf{Objective}				
	\\
	\midrule
Response time						&
0.50										&
ms											&
25	
\\
Greenability							&
0.20										&
kWh										&
25 
\\
Operational cost					&
0.20										&
\$												&
50		
\\
	\midrule
Frequency of adaptation			&
-														&
cycles											&
-	
\\
Adaptation overhead			&
-												&
ms											&
-		
\\
Resources overshoot			&
-													&
PMs, VMs									&
-		
\\
\bottomrule
\end{tabularx}
\end{table}

We define the catalogue of architectural tactics to fulfil the stability attributes subject to consideration. Table \ref{tbl_QualityTactics} lists the tactics and their definitions. We base this work on the description tactics by Bass et al. \cite{Bass2012}. The tactics include: (i) horizontal scaling (increasing/decreasing the number of physical machines), (ii) vertical scaling (increasing/decreasing the number of virtual machines or their CPU capacities), (iii) virtual machines consolidation (running the virtual machines on less number of physical machines for energy savings), (iv) concurrency (by processing different streams of events on different threads or by creating additional threads to process different sets of activities), (v) dynamic priority scheduling (scheduling policy is implemented, where the scheduler handles requests according to a scheduling policy), and (vi) energy monitoring (providing detailed energy consumption information). Adaptation rules, embedded in the stimulus-awareness component, are defined as such tactics related with stability attributes. Adaptation rules are illustrated in Table \ref{tbl_AdaptationRules}.

\begin{landscape}
\begin{table}[!h]
\caption{Adaptation tactics and their definitions}
\label{tbl_QualityTactics}
\center
\footnotesize
\begin{tabularx}{1.40\textwidth}
{
l
>{\raggedright}p{0.20\textwidth} 
>{\raggedright}p{0.30\textwidth} 
>{\raggedright}p{0.15\textwidth} 
>{\raggedright}p{0.25\textwidth} 
>{\raggedright\arraybackslash}p{0.30\textwidth} 
}
	\toprule
	\textbf{No.}						&
	\textbf{Tactic}					&	
	\textbf{Description}		&	
	\textbf{Object}				&
	\textbf{Limits}					&	
	\textbf{Variations}			
	\\
	\midrule
1											&
Vertical scaling 				&
increasing the number of virtual machines (VMs) or their CPU capacities							&
VMs									&
maximum CPU capacity of hosts running in the datacenter		&
+1, 2, 3,... VMs or increase the CPU capacity of running VMs	
\\
2											&
Vertical de-scaling 				&
decreasing the number of virtual machines (VMs) or their CPU capacities									&
VMs											&
minimum one running VM		&
+1, 2, 3,... VMs						
\\
3											&
Horizontal scaling			&
increasing the number of running hosts					&
Hosts									&
maximum number of hosts in the datacenter			&
+1, 2, 3,... hosts				
\\
4											&
Horizontal de-scaling					&
decreasing the number of running hosts					&
Hosts													&
minimum one running host 			&
-1, 2, 3,... hosts								
\\
5											&
VMs consolidation			&
shut down hosts running least number of VMs and migrate their VMs to other hosts									&
Hosts, VMs												&
minimum one running host and one VM 			&	 
-1, 2, 3,... hosts										
\\
6															&
Concurrency 									&
processing different streams of events on different threads or by creating additional threads to process different sets of activities		&
datacenter scheduler					&
maximum CPU capacity of hosts running in the datacenter					&
single, multiple threads					
\\
7												&
Dynamic scheduling			&
scheduling policy is implemented, where the scheduler handles requests according to a scheduling policy						&
datacenter scheduler						&
maximum number of running hosts and VMs					&
earliest deadline first scheduling, least slack time scheduling, single queueing, multiple queueing, multiple dynamic queueing				
\\
\bottomrule
\end{tabularx}
\end{table}
\end{landscape}

\begin{table}[!h]
\caption{Adaptation Rules}
\label{tbl_AdaptationRules}
\center
\footnotesize
\begin{tabular}
{ lll }
	\toprule
	\textbf{Tactic}												&	
	\textbf{Related Quality Attributes}		&
	\textbf{Priority}
	\\
	\midrule
Dynamic scheduling						&
response time									&
1
\\
Conucrrency 									&
response time									&
2
\\
Vertical scaling 									&
response time					&
3
\\
Horizontal scaling							&
response time				&
4
\\
VMs consolidation												&
energy consumption			&
1
\\
Vertical de-scaling 												&
operational cost, energy consumption				&
2
\\
Horizontal de-scaling										&
operational cost, energy consumption			&
3
\\
	\bottomrule
\end{tabular}
\end{table}

We used benchmarks to stress the architecture with highly frequent changing demand and observe stability goals. To simulate runtime dynamics, we used the \textit{RUBiS} benchmark \cite{RUBiS} and the \textit{World Cup 1998} trend \cite{WorldCup98} in our experiments. The \textit{RUBiS} benchmark \cite{RUBiS} is an online auction application defining different services categorised in two workload patterns: the browsing pattern (read-only services, e.g. BrowseCategories), and the bidding pattern (read and write intensive services, e.g. PutBid, RegisterItem, RegisterUser). For fitting the simulation parameters, we mapped the different services of the \textit{RUBiS} benchmark into Million Instructions Per Second (MIPS), as listed in Table \ref{tbl_experimentsServices}. To simulate a realistic workload within the capacity of our testbed, we varied the number of requests proportionally according to the \textit{World Cup 1998} workload trend \cite{WorldCup98}. We compressed the trend in a way that the fluctuation of one day (=86400 sec) in the trend corresponds to one time instance of 864 seconds in our experiments. This setup can generate up to 700 parallel requests during one time instance, which is large enough to challenge stability. 

\begin{table}[!h]
\caption{Types of service requests}
\label{tbl_experimentsServices}
\center
\footnotesize
\begin{tabularx}{0.95\textwidth}
{
		>{\raggedright}p{0.27\textwidth} 
		l
		l
		l
}
\toprule
{\textbf{Service Pattern}} 		& 
{\textbf{S\#}} 								& 
{\textbf{Service Type}} 			& 
{\textbf{Required MIPS}}
\\
\midrule
browsing only			&
1									&
read-only					&
10,000  
\\
\hline
bidding only				&
2									&
read and write			&
20,000  
\\
\hline
\multirow{3}{*}{\parbox{3cm}{mixed with adjustable composition of the two service patterns}}															&
3																											&
70\% browsing, 30\% bidding												&
12,000  
\\
																											&
4																											&
50\% browsing, 50\% bidding												&
15,000  
\\
																											&
5																											&
30\% browsing, 70\% bidding												&
17,000  
\\
\bottomrule
\end{tabularx}
\end{table}

The initial deployment of the experiments is 10 running hosts IBM x3550 server, each with the configuration of 2 x Xeon X5675 3067 MHz, 6 cores and 256 GB RAM. The frequency of the servers' CPUs is mapped onto MIPS ratings: 3067 MIPS each core \cite{Beloglazov2012} and their energy consumption is calculated using power models of \cite{Beloglazov2012}. The maximum capacity of the architecture is 1000 hosts. The characteristics of the virtual machines (VMs) types correspond to the latest generation of General Purpose Amazon EC2 Instances \cite{AmazonEC2}. In particular, we use the m4.large (2 core vCPU 2.4 GHz, 8 GB RAM), m4.xlarge (4 core vCPU 2.4 GHz, 16 GB RAM), and m4.2xlarge (8 core vCPU 2.4 GHz, 32 GB RAM) instances. The operational cost of different VMs types is 0.1, 0.2 and 0.4 \$/hour respectively. Initially, the VMs are allocated according to the resource requirements of the VM types. However, VMs utilise less resources according to the workload data during runtime, creating opportunities for dynamic consolidation. The initial deployment of the experiment is shown in Table \ref{tbl_experimentsConfiguration}.

\begin{table}[!h]
\caption{Initial deployments of the experimental evaluation}
\label{tbl_experimentsConfiguration}
\center
\footnotesize
\begin{tabularx}{0.70\textwidth}
{
l
>{\raggedright\arraybackslash}p{0.50\textwidth} 
}
\toprule
{\textbf{Configuration}} 		& 	
{\textbf{}}
\\
\midrule
No. of hosts	&
10 running (max. 1000)
\\
\hline
Host type		&
IBM x3550 server
\\
\hline
Host Specs	&
2 x Xeon X5675 3067 MHz, \newline 6 cores, 256 GB RAM
\\
\hline
No. of VMs	&	
15
\\
\hline
VMs type		&
General Purpose Amazon EC2 Instances \newline 
m4.large, m4.xlarge, m4.2xlarge	
\\
\hline
VMs Capacity	&	
5 x 2 core CPU 8 GB RAM,	\newline 
5 x 4 core CPU 16 GB RAM,	\newline 
5 x 8 core CPU 32 GB RAM
\\
\bottomrule
\end{tabularx}
\end{table}

\subsection{Results of Stability Attributes}
\label{sec_evaluation_results}
We report, first, on the average of stability attributes for each service type of 30 runs. We examined stability attributes at each time interval of 864 seconds. More specifically, we run the entire workload for each service type and measured the stability attributes when using each self-aware capability compared to self-adaptive architecture. The implemented self-adaptive architecture is a self-adaptive MAPE architecture \cite{Maurer2011}.

\begin{figure}[!h]
\centering
\includegraphics[width=0.75\textwidth]{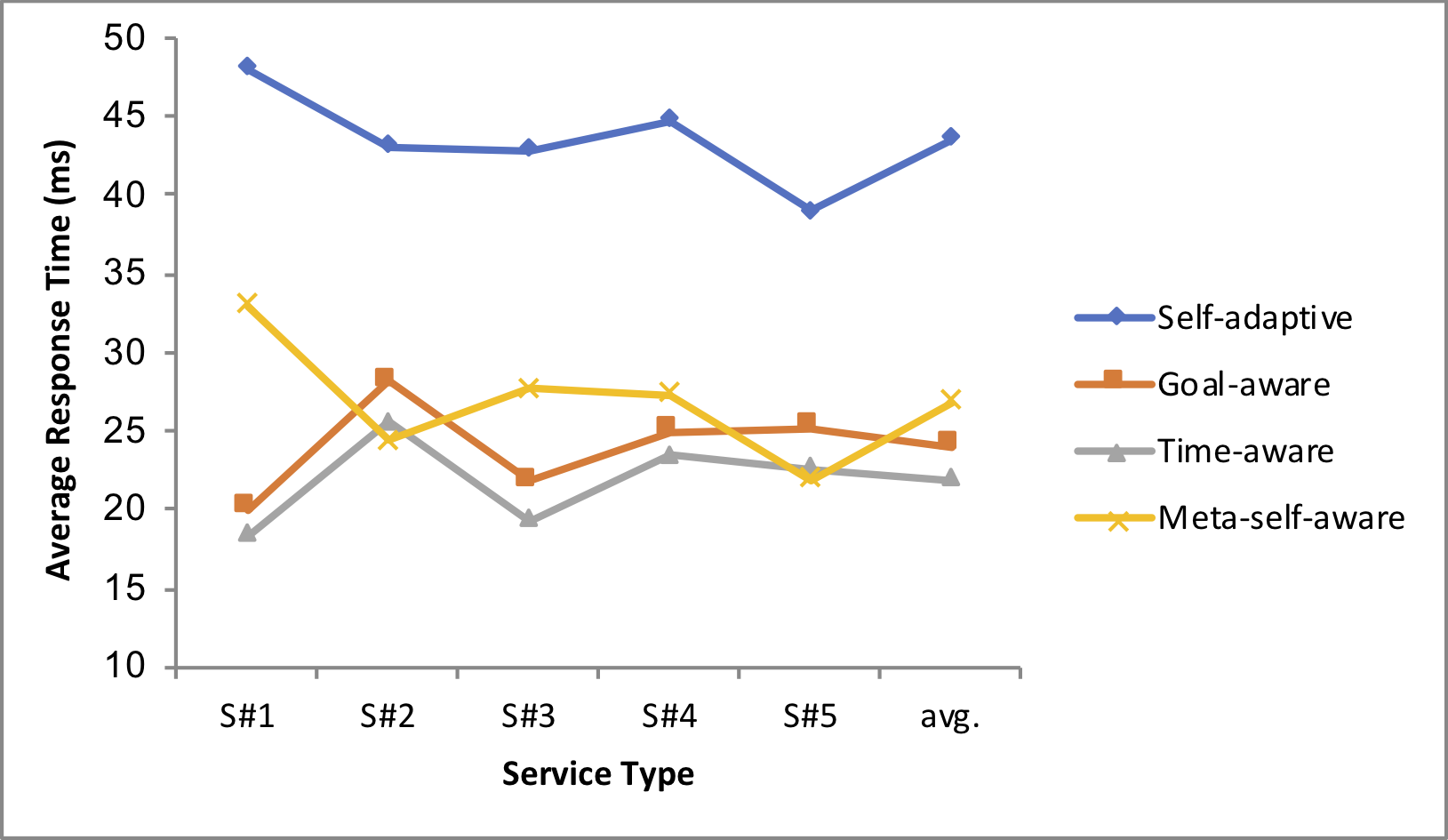}
\caption{Average Response Time (ms)}
\label{graph_resultsResponseTime}
\end{figure}

\begin{figure}[!h]
\centering
\includegraphics[width=0.75\textwidth]{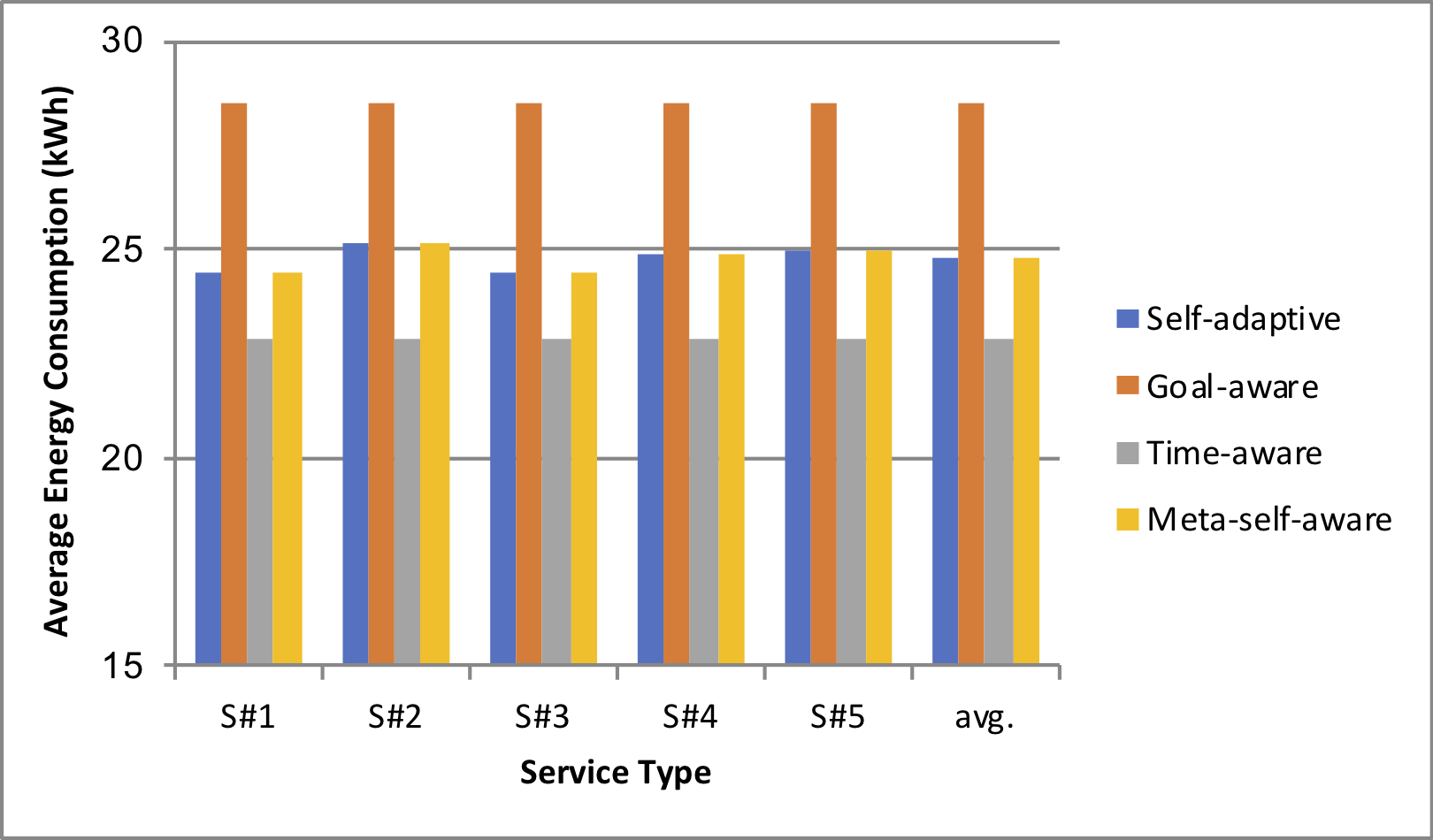}
\caption{Average Energy Consumption (kWh)}
\label{graph_resultsEnergy}
\end{figure}

\begin{figure}[!h]
\centering
\includegraphics[width=0.75\textwidth]{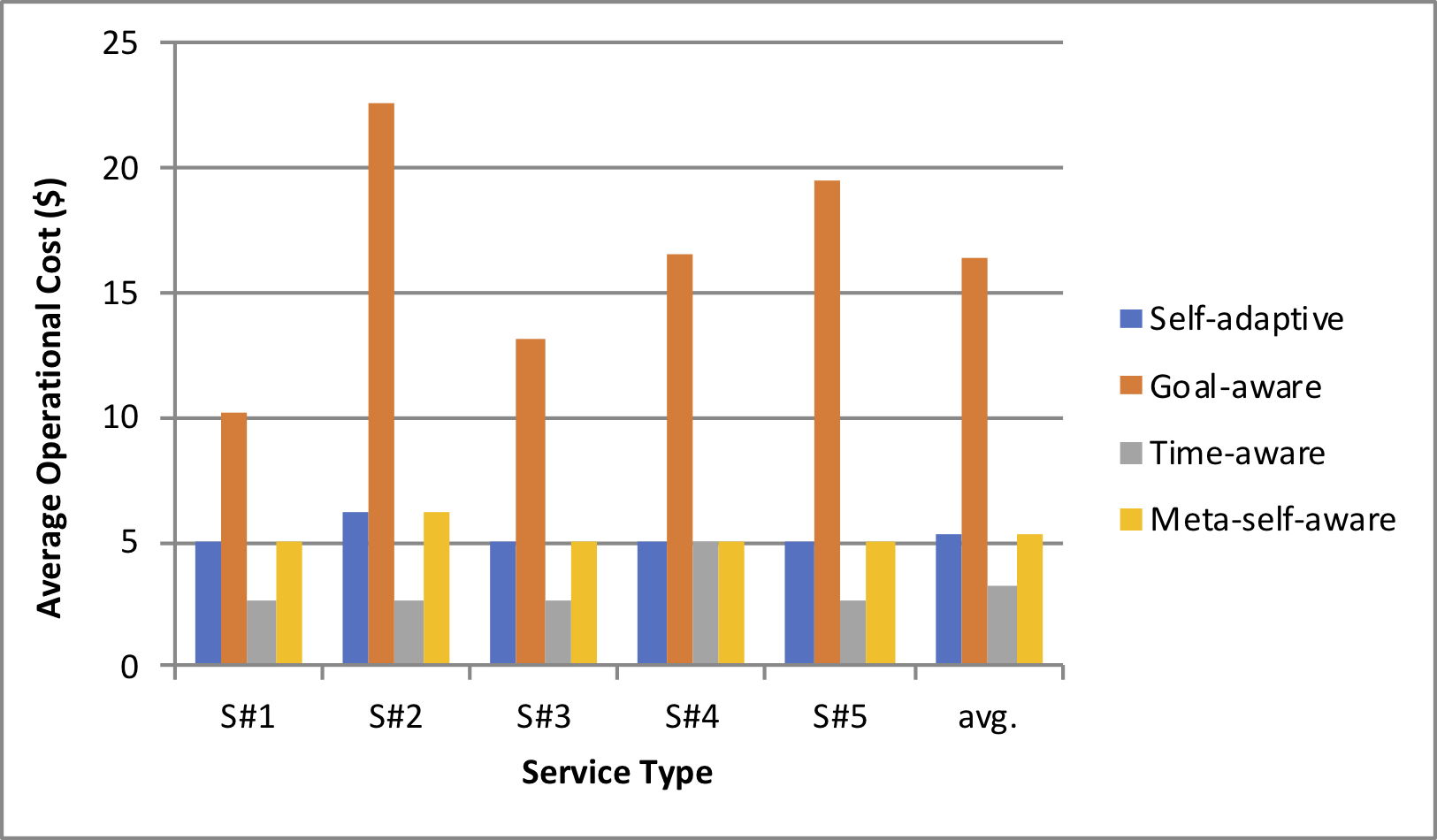}
\caption{Average Operational Cost (\$)}
\label{graph_resultsCost}
\end{figure}

The average of response time, energy consumption and operational costs are depicted in Figure \ref{graph_resultsResponseTime}, \ref{graph_resultsEnergy} and \ref{graph_resultsCost} respectively. On average, the self-awareness capabilities outperformed the self-adaptive one in keeping response time (highest priority) within stability objective. As shown in Figure \ref{graph_resultsResponseTime}, time-awareness achieved the best response time for all service types. Meanwhile, meta-self-awareness was capable to achieve the best performance for service type 2 and 5 which require the higher computational resources.

Regarding energy consumption, while all awareness algorithms succeeded in maintaining the energy consumption stability objective, time-awareness has consumed less energy reflecting the minimal number of PMs running (resources overshoot). This is due to performing adaptations that are capable to keep stability goals for longer periods. Meanwhile, goal-awareness used the highest number of hosts, due to more frequent adaptation (frequent shut-down and re-run of hosts) to keep stability goals. Meta-self-awareness was capable to maintain the trade-offs between energy consumption and response time. 

Similar to energy consumption, operational costs (reflecting the number of VMs running) was better achieved by time-awareness, followed by meta-self-awareness. Goal-awareness has the highest cost, even though within stability objective.

\subsection{Evaluation of Adaptation Properties}
\label{sec_evaluation_adaptation}
\begin{figure}[!h]
\centering
\includegraphics[width=0.75\textwidth]{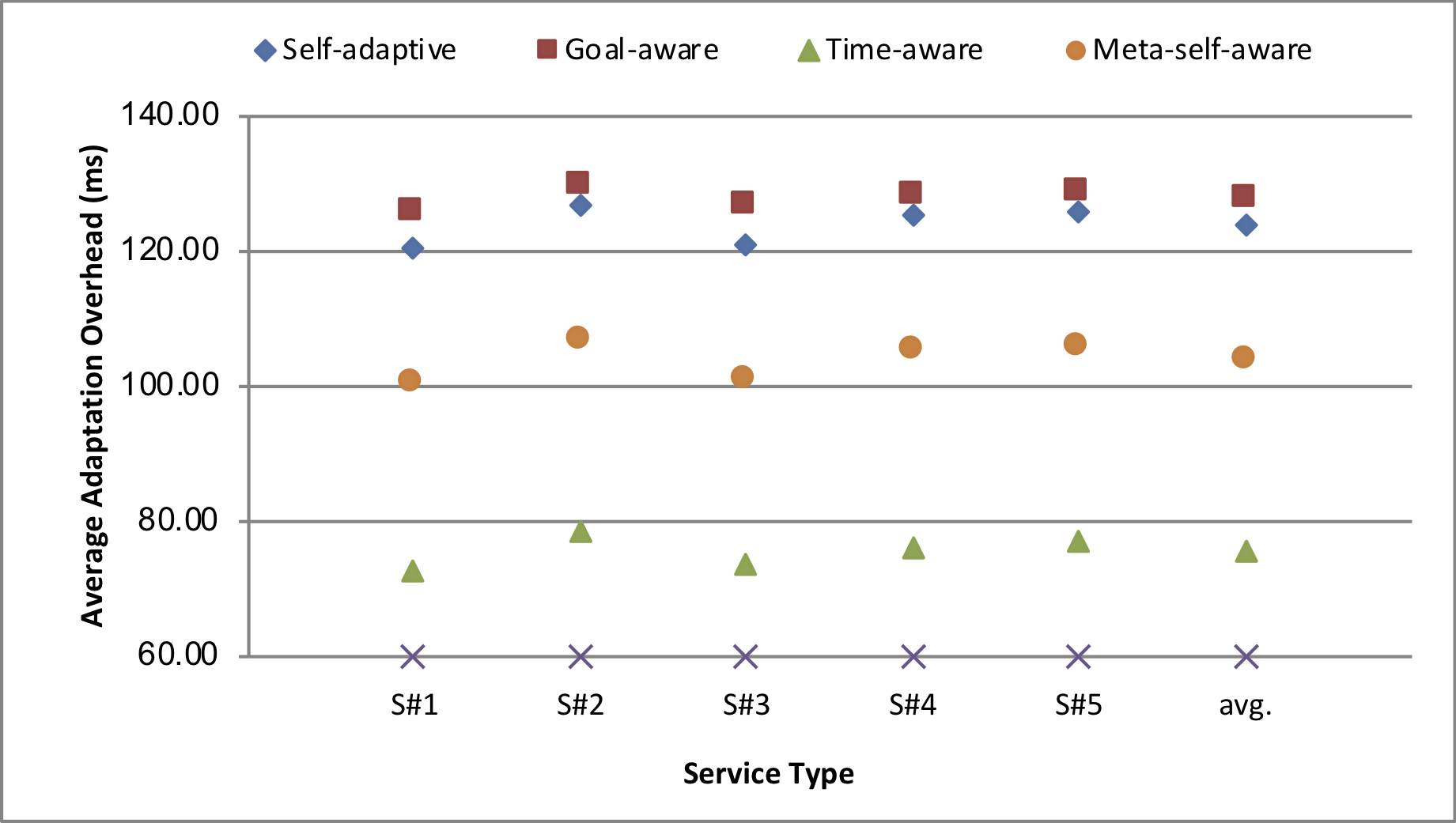}
\caption{Frequency of Adaptation (no. of adaptation cycles)}
\label{graph_resultsAdaptationFrequency}
\end{figure}

Given the direct impact of the frequency of adaptation on architectural stability, we evaluate the number of adaptation cycles taken by each capability. As shown in Figure \ref{graph_resultsAdaptationFrequency}, time-awareness performed the least number of adaptation cycles, followed by meta-self-awareness. Meanwhile, goal-awareness is higher and close to self-adaptive, but achieved better response time than self-adaptive.

\begin{figure}[!h]
\centering
\includegraphics[width=0.75\textwidth]{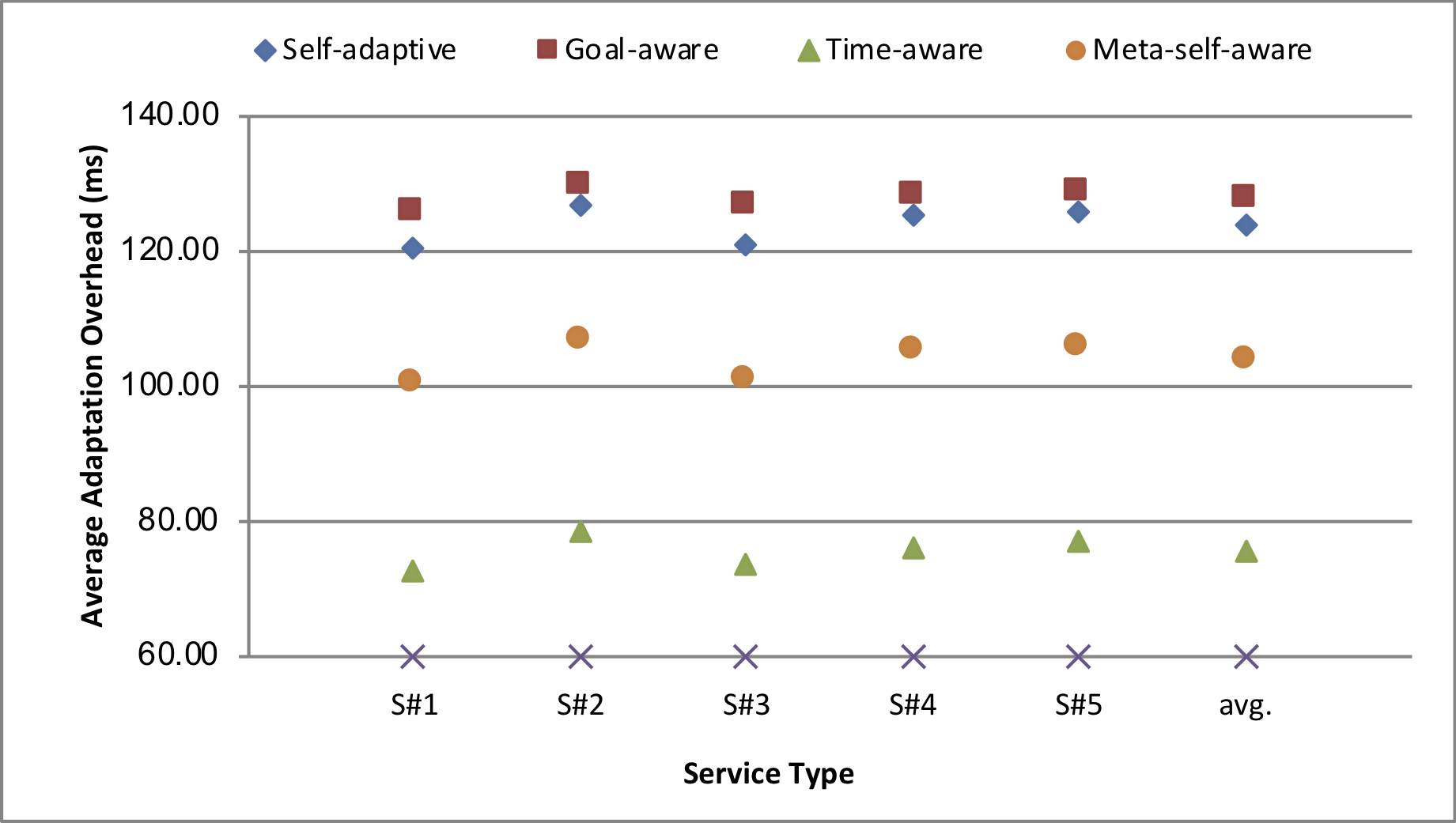}
\caption{Adaptation Overhead (sec)}
\label{graph_resultsAdaptationOverhead}
\end{figure}

We evaluate the adaptation overhead by calculating the total time spent by the architecture in the adaptation process. Figure \ref{graph_resultsAdaptationOverhead} shows the overhead of each service type and their average. As goal-awareness performs pro-active adaptations for keeping stability gaols, its overhead is higher than self-adaptive (127.78 versus 123.78 sec on average), which obviously resulted in better response time. Meta-self-awareness is on average of all capabilities (103.78 sec on average), while time-awareness has achieved the lower overhead (75.42 sec on average).

\subsection{Discussion}
\label{sec_evaluation_discussion}
The proposed framework with different self-awareness capabilities have successfully achieved stability in terms of quality attributes and adaptation properties under runtime changing workload. Evaluating the features of the proposed framework is summarised as follows. First, different self-awareness principles were capable to successfully achieve stability attributes, combining quality attributes subject to stability and quality of adaptation. The generic framework allowed featuring different combinations of self-awareness capabilities for reasoning about long-term stability using machine learning and stochastic games techniques. Further, these reasoning techniques could be extended easily. Also, different stability goals could be easily configured, and other reasoning techniques could be employed. The proposed framework and architecture are generic for using one single self-aware capability, as well as switching between different capabilities during runtime. 

Generally, the proposed approaches have proven feasibility in reasoning about stability during runtime, where the implemented components tend to make more intelligent adaptation actions. The quantitative evaluation has proven their ability to efficiently reason about stability, avoid unnecessary frequent adaptations and minimise adaptation overhead and resources overshoot.

\section{Threats to Validity}
\label{sec_threats}
The potential threats to validity of the proposed work are:

\begin{itemize}
\item Subjectivity might be considered a threat to validity in setting the stability attributes, as it was conducted based on the authors' background and knowledge. Our mitigation strategy for this issue is to base the evaluation case on previous work of \cite{Chen2014a} \cite{Salama2015} \cite{Salama2016} \cite{Salama2017b}, this makes us believe that the case study is practical and reflects the nature of cloud-based software systems.

\item Another threat to validity of our evaluation lies in the fact that the approach was evaluated using one case. Yet, the dynamics presented in cloud architectures is an appropriate case study representing dynamics of modern software systems, and we plan to conduct other case studies in industrial contexts and different business segments. 

\item Experiments were conducted in a controlled environment and have not considered the real-life scenario of switching between different service patterns and changing stability goals during runtime for different end-users. Given the use of a real-world workload trend and the \textit{RUBiS} benchmark, we consider that our experiments have given good enough indication and approximation of likely scenarios in a practical setting. Also, we have chosen the stability goals thresholds purely based on our observations, e.g. response time not exceeding 25 ms. Yet, these goals have been tested and proved to be challenging. 

\end{itemize}

\section{Related Work}
\label{sec_relatedWork}
In this section, we discuss related work to goals modelling (section \ref{sec_relatedWork_goals}), learning for self-adaptation (section \ref{sec_relatedWork_learning}) and trade-offs management in self-adaptive systems (section \ref{sec_relatedWork_tradeoffs}).

\subsection{Goals Modelling}
\label{sec_relatedWork_goals}
Seminal works related to runtime requirements modelling are ``models@run.time'' and ``self-explanation''. 

Models@run.time rethinks adaptation mechanisms in a self-adaptive system by leveraging on model-driven engineering approaches to the applicability at runtime \cite{Blair2009}. This approach supports requirements monitoring and control, by dynamically observing the runtime behaviour of the system during execution. Models@run.time can interleave and support runtime requirements, where requirements and goals can be observed during execution by maintaining a model of the requirements in conjunction to its realisation space. The aim is to monitor requirements satisfaction and provide support for unanticipated runtime changes by tailoring the design and/or invoking adaptation decisions which best satisfy the requirements. Meanwhile, authors in \cite{Lamsweerde2003} proposed a goal-oriented approach for systematically building architecture design from system goals.

In the context of self-adaptive systems, self-explanation was introduced to adaptive systems to offer interpretation of how a system is meeting its requirements, using goal-based requirements models \cite{Welsh2014}. Self-explanation focused mainly on explaining the self-adaptive behaviour of the running system, in terms of satisfaction of its requirements, so that developers can understand the observed adaptation behaviour and garner confidence to its stakeholders. Authors in \cite{Cavalcante2015} have theoretically revisited goal-oriented models for self-aware systems-of-systems. Goal models were also introduced as runtime entities in adaptive systems \cite{Goldsby2008} and context-aware systems \cite{Vrbaski2012}.

Though, there has been growing research in runtime requirements engineering in the context of self-adaptive software systems, yet these models and approaches have limitations in enabling the newly emerged self-properties, i.e. self-awareness and self-expression. To the best of our knowledge, there is no research that tackled goals modelling for self-aware and self-expressive software systems, as well realising the symbiotic relation between both.

\subsection{Learning for Self-Adaptation}
\label{sec_relatedWork_learning}
Learning for self-adaptation has been studied by a number of researchers using different learning techniques for different purposes. For instance, a learning approach for engineering feature-oriented self-adaptive systems has been proposed in \cite{Esfahani2013c}, learning revised models for planning self-adaptive systems \cite{Sykes2013}, modelling self-adaptive systems with Learning Petri Nets \cite{Ding2016}, and handling uncertainty using self-learning fuzzy neural networks \cite{Han2016}, 

Focusing on the dynamic learning behaviour during runtime operation of adaptive systems, Yerramalla et al. \cite{Yerramalla2005} have proposed a stability monitoring approach based on Lyapunov functions for detecting unstable learning behaviour, and mathematically analysed stability to guarantee that the runtime learning converges to a stable state within reasonable time depending on the application. Yet, quality of adaptation has not been considered in the stability behaviour. A reinforcement learning-based approach has also been proposed for planning architecture-based self-management \cite{Dongsun2009}. Meanwhile, the behavioural stability aspect we are seeking has not been learnt online.

\subsection{Trade-offs Management}
\label{sec_relatedWork_tradeoffs}
Research has encountered many efforts for managing architectural trade-offs and the field has attracted a wide range of researchers and practitioners. Seminal works for trade-offs management include Architecture Tradeoff Analysis Method (ATAM) \cite{Kazman1998}, Cost Benefit Analysis Method (CBAM) \cite{Nord2003}, PerOpteryx \cite{Koziolek2011}, the work of Kazman et al. \cite{Kazman1994a} and the Quality-attribute-based Economic Valuation of Architectural Patterns \cite{Ozkaya2007}. Despite the maturity of research in evaluating and analysing architecture trade-offs, self-adaptive architectures call for special treatment, since self-adaptation has been primarily driven by the need to achieve and maintain quality attributes in the face of the continuously changing requirements and uncertain demand at runtime, as a result of operating in dynamic and uncertain contexts.

In our prior work \cite{Salama2017a}, we have systematically surveyed the literature related to trade-offs management for self-adaptive architectures. We differentiated between approaches for design-time and runtime. By design-time, we mean trade-offs management is considered while evaluating the architectural design alternatives and making architectural decisions. The runtime is meant to be managing trade-offs while the system is operating and the change requests are implemented.

Our findings show some attention given for explicit consideration of trade-offs management at runtime. Examples include \cite{Teich2009} \cite{Mirandola2010} \cite{Menasce2011} \cite{Perez-Palacin2012} \cite{Peng2012} \cite{Shen2012}. But analysing the research landscape \cite{Salama2017a}, our observation is that there is an adoption for the general ``self-adaptivity'' property without a discrete specialisation on self-* properties. The generality also applies to the quality attributes considered in trade-offs management. When considering certain qualities, they tend to be limited to two or three attributes, as explicit examples. As a general conclusion, the current work tends to be a solution for trade-offs management that act on trade-offs, not fundamental work that changes the architectural self-adaptivity. Although the studies found have provided much that is useful in contributing towards self-adaptive architectures, it has not yet resolved some of the general and fundamental issues in order to provide a comprehensive, systematic and integrated approach for runtime support for change and uncertainty while managing trade-offs.

With respect to considering multiple quality concerns, Cheng et al. \cite{Cheng2012} have presented a language for expressing adaptation strategies to calculate the best strategy for decision-making to be carried out by system administrators. Though, this work has considered multiple QoS objectives, and represented uncertainties in adaptation outcome, it is useful only for human operators use, not autonomous use during runtime. The work of Camara et al. \cite{Camara2014c} \cite{Camara2015} has employed stochastic games for proactive adaptation, to balance between the cost and benefits of a proactive approach for adaptation. Meanwhile, self-adaptive architectures need to ensure the provision of multiple quality attributes. This also requires considering the quality of adaptation \cite{Villegas2011}, as well as the cost and overhead of adaptation. Yet, our work considers architectural stability in terms of quality provision and quality of adaptation, using stochastic games with multi-objectives queries and long-run rewards.

\section{Conclusion and Future Work}
\label{sec_conclusion}
In this paper, we presented self-awareness techniques for reasoning about architectural behavioural stability during runtime. We presented runtime goals model for managing stability goals using self-awareness. We also implemented an online learning technique for reasoning about stability on the long-run while learning from historical information. Trade-offs management between different stability attributes is managed using model verification of stochastic games. Using the case of cloud architecture, quantitative experiments have proven enhancements in achieving stability and quality of adaptation when using different self-awareness techniques. 

Our future work will focus on switching between different self-awareness techniques during runtime, to achieve better results. We also plan to conduct other evaluation cases for our framework, such as scientific workflows.

\section*{Acknowledgments}
We would like to thank Abdessalam Elhabbash for useful discussions on self-awareness, Rafik Salama for online learning inspiration, and David Parker for the support on PRISM-games.

%\section*{References}
\bibliographystyle{plain}
\bibliography{JStabilityReasoning-bib}

\end{document}